\documentclass{emulateapj}
\usepackage{apjfonts}

\def\asca{{\it ASCA\/}}

\def\chandra{{\it Chandra\/}}

\def\genx{{\it Generation-X\/}}

\def\jwst{{\it {\it JWST}\/}}
\def\ixo{{\it {\it IXO}\/}}
\def\wfxt{{\it {\it WFXT}\/}}
\def\hst{{\it {\it HST}\/}}

\def\iras{{\it IRAS\/}}

\def\spitzer{{\it Spitzer\/}}
\def\galex{{\it GALEX\/}}

\def\xray{\hbox{X-ray}}

\def\etal{{et\,al.}}

\def\ltsima{$\; \buildrel < \over \sim \;$}
\def\simlt{\lower.5ex\hbox{\ltsima}}
\def\gtsima{$\; \buildrel > \over \sim \;$}
\def\simgt{\lower.5ex\hbox{\gtsima}}
\def\kms{\ifmmode{~{\rm km~s^{-1}}}\else{~km s$^{-1}$}\fi}
\def\lsim{\lower0.3em\hbox{$\,\buildrel <\over\sim\,$}}
\def\gsim{\lower0.3em\hbox{$\,\buildrel >\over\sim\,$}}

\def\h2{H$_2$}

\def\xlum{ergs~s$^{-1}$}

\def\arcsec{\mbox{$^{\prime\prime}$}}
\def\arcmin{\mbox{$^\prime$}}
\def\sfr{$M_{\odot}$ yr$^{-1}$}

\def\aap{A\&A}
\def\apj{ApJ}
\def\apjl{ApJL}
\def\apjs{ApJS}
\def\aj{AJ}
\def\mnras{MNRAS}
\def\araa{ARA\&A}

\begin{document}

\shortauthors{LEHMER ET AL.}
\shorttitle{X-ray Emission from LIRGs}

%
%%%%%%%%%%%%%%%%%%%%%%%%%%%%%%%%%%%%%%%%%%%%%%%%%%%%%%%%%%%%%%%%%%%%%%%%
\title{A {\itshape Chandra} Perspective on Galaxy-Wide X-ray Binary Emission
and its Correlation with Star Formation Rate and Stellar Mass: New
Results from Luminous Infrared Galaxies.}
%%%%%%%%%%%%%%%%%%%%%%%%%%%%%%%%%%%%%%%%%%%%%%%%%%%%%%%%%%%%%%%%%%%%%%%%
%

\author{
B.~D.~Lehmer,\altaffilmark{1,2}
D.~M.~Alexander,\altaffilmark{3}
F.~E.~Bauer,\altaffilmark{4,5}
W.~N.~Brandt,\altaffilmark{6,7}
A.~D.~Goulding,\altaffilmark{3} 
L.~P.~Jenkins,\altaffilmark{1,2}
A.~Ptak,\altaffilmark{2} \&
T.~P.~Roberts\altaffilmark{3}
}
\altaffiltext{1}{The Johns Hopkins University, Homewood Campus, Baltimore, MD 21218, USA}
\altaffiltext{2}{NASA Goddard Space Flight Centre, Code 662, Greenbelt, MD 20771, USA} 
\altaffiltext{3}{Department of Physics, University of Durham, South Road, Durham, DH1 3LE, UK} 
\altaffiltext{4}{Pontificia Universidad Catolica de Chile, Departamento de Astronomia y Astrofisica, Casilla 306, Santiago 22, Chile} 
\altaffiltext{5}{Space Science Institute, 4750 Walnut Street, Suite
205, Boulder, Colorado 80301}
\altaffiltext{6}{Department of Astronomy \& Astrophysics, 525 Davey Lab,
The Pennsylvania State University, University Park, PA 16802, USA} 
\altaffiltext{7}{Institute for Gravitation and the Cosmos, The Pennsylvania State University, University Park, PA 16802, USA}

%
%%%%%%%%%%%%%%%%%%%%%%%%%%%%%%%%%%%%%%%%%%%%%%%%%%%%%%%%%%%%%%%%%%%%%%%%
\begin{abstract}
%%%%%%%%%%%%%%%%%%%%%%%%%%%%%%%%%%%%%%%%%%%%%%%%%%%%%%%%%%%%%%%%%%%%%%%%
%

\noindent We present new \chandra\ observations that complete a sample of
seventeen (17) luminous infrared galaxies (LIRGs) with $D < 60$~Mpc and low
Galactic column densities of $N_{\rm H} \simlt 5 \times 10^{20}$~cm$^{-2}$.
The LIRGs in our sample have total infrared (\hbox{8--1000$\mu$m}) luminosities
in the range of $L_{\rm IR} \approx$~(1--8)~$\times$~$10^{11}$~$L_\odot$.  The
high-resolution imaging and \xray\ spectral information from our \chandra\
observations allow us to measure separately \xray\ contributions from active
galactic nuclei (AGNs) and normal galaxy processes (e.g., \xray\ binaries and
hot gas).  We utilized total infrared plus UV luminosities to estimate
star-formation rates (SFRs) and $K$-band luminosities and optical colors to
estimate stellar masses ($M_\star$) for the sample.  Under the assumption that
the galaxy-wide \hbox{2--10~keV} luminosity ($L_{\rm HX}^{\rm gal}$) traces the
combined emission from high mass \xray\ binaries (HMXBs) and low mass \xray\
binaries (LMXBs), and that the power output from these components are linearly
correlated with SFR and $M_\star$, respectively, we constrain the relation
$L_{\rm HX}^{\rm gal} = \alpha M_\star + \beta {\rm SFR}$.  To achieve this, we
construct a \chandra-based data set composed of our new LIRG sample combined
with additional samples of less actively star-forming normal galaxies and more
powerful LIRGs and ultraluminous infrared galaxies (ULIRGs) from the
literature.  Using these data, we measure best-fit values of $\alpha = (9.05
\pm 0.37) \times 10^{28}$~ergs~s$^{-1}$~$M_\odot^{-1}$ and $\beta = (1.62 \pm
0.22) \times 10^{39}$~ergs~s$^{-1}$~($M_\odot$~yr$^{-1}$)$^{-1}$.  This scaling
provides a more physically meaningful estimate of $L_{\rm HX}^{\rm gal}$, with
\hbox{$\approx$0.1--0.2~dex} less scatter, than a direct linear scaling with
SFR.  Our results suggest that HMXBs dominate the galaxy-wide \xray\ emission
for galaxies with SFR/$M_\star$~$\simgt 5.9\times 10^{-11}$~yr$^{-1}$, a factor
of $\approx$2.9 times lower than previous estimates.  We find that several of
the most powerful LIRGs and ULIRGs, with SFR/$M_\star \simgt
10^{-9}$~yr$^{-1}$, appear to be \xray\ underluminous with respect to our
best-fit relation.  We argue that these galaxies are likely to contain \xray\
binaries residing in compact star-forming regions that are buried under thick
galactic columns large enough to attenuate emission in the \hbox{2--10~keV}
band ($N_{\rm H} \simgt 10^{23}$~cm$^{-2}$).

%
%%%%%%%%%%%%%%%%%%%%%%%%%%%%%%%%%%%%%%%%%%%%%%%%%%%%%%%%%%%%%%%%%%%%%%%%
\end{abstract}
%%%%%%%%%%%%%%%%%%%%%%%%%%%%%%%%%%%%%%%%%%%%%%%%%%%%%%%%%%%%%%%%%%%%%%%%
%


%
%%%%%%%%%%%%%%%%%%%%%%%%%%%%%%%%%%%%%%%%%%%%%%%%%%%%%%%%%%%%%%%%%%%%%%%%
\section{Introduction}
%%%%%%%%%%%%%%%%%%%%%%%%%%%%%%%%%%%%%%%%%%%%%%%%%%%%%%%%%%%%%%%%%%%%%%%%
%

The most energetic nearby star-forming galaxies are luminous infrared galaxies
(LIRGs), which are classified as having total infrared (\hbox{8--1000$\mu$m})
luminosities greater than $10^{11}$~$L_{\odot}$ (where $L_{\odot} = 3.9 \times
10^{33}$~ergs~s$^{-1}$ is the bolometric luminosity of the Sun).  The total
infrared emission from most LIRGs comprises $\simgt$50--90\% of their galactic
bolometric power output and is thought to arise primarily from dust
reprocessing of obscured UV light that originates from underlying star-forming
regions (see Sanders \& Mirabel~1996 for a review).  Therefore, the total LIRG
infrared power provides an effective measure of the galactic star-formation
rate (SFR; e.g., Kennicutt~1998) and has been used extensively to select
galaxies with active star-formation (e.g., Sanders \etal\ 2003).  

%%%%%%%%%%%%%%%%%%%%%%%%%%%%%%%%%%%%%%%%%%%%%%%%%%%%%%%%%%%%%%%%%%%%%%%%%%%%%%%%%%
% Table 1
%%%%%%%%%%%%%%%%%%%%%%%%%%%%%%%%%%%%%%%%%%%%%%%%%%%%%%%%%%%%%%%%%%%%%%%%%%%%%%%%%%
\begin{table*}
{\tiny
\begin{center}
\caption{LIRG Sample List and Basic Properties}
\begin{tabular}{lccccccccccccc}
\hline\hline
 &  \multicolumn{5}{c}{\sc $K_s$-Band Source Characteristics} & & & & & & & \\
 &  \multicolumn{5}{c}{\rule{2in}{0.01in}} & & & & & & & \\
  & $\alpha_{\rm J2000}$ & $\delta_{\rm J2000}$ & $a$ & $b$ & PA & $D_{\rm L}$ & $N_{\rm H}$ & $\log L_{\rm IR}$ & SFR & $\log M_\star$ & Optical & Optical & Class \\
\multicolumn{1}{c}{\sc Source Name} & (hrs) & (deg) & (arcmin) & (arcmin) & (deg) & (Mpc) & ($10^{20}$~cm$^{-2}$) & ($L_{\odot}$) & ($M_\odot$/yr) & ($M_\odot$) & Morphology & Class & Reference \\
\multicolumn{1}{c}{(1)} & (2) & (3) & (4) & (5) & (6) & (7) & (8) & (9) & (10) & (11) & (12) & (13) & (14) \\
\hline\hline
                                                       NGC~1068\ldots\ldots &      02 42 41 &   $-$00 00 48 &   2.42 &   1.99 &  35 & 13.75 &  4.06 & 11.27 &    18.50 &  11.02 &           (R)SA(rs)b &                         Sy~2 &            1,2,3,4,5 \\
                                                           NGC~1365\dotfill &      03 33 36 &   $-$36 08 25 &   4.60 &   3.45 &  50 & 18.03 &  1.36 & 11.00 &     9.97 &  11.01 &               SB(s)b &                         Sy~1 &                    1 \\
                                                           NGC~7552\dotfill &      23 16 11 &   $-$42 35 05 &   2.21 &   1.30 &  95 & 21.55 &  1.95 & 11.03 &    10.67 &  10.70 &          (R')SB(s)ab &           H~{\scriptsize II} &                1,4,5 \\
                                                           NGC~4418\dotfill &      12 26 55 &   $-$00 52 39 &   0.83 &   0.45 &  55 & 32.12 &  2.04 & 11.08 &    11.79 &  10.04 &          (R')SAB(s)a &             {\sc LINER}/Sy~2 &                5,6,7 \\
                                                           NGC~4194\dotfill &      12 14 09 &     +54 31 37 &   0.64 &   0.44 & 170 & 40.67 &  1.56 & 11.06 &    11.26 &  10.40 &              IBm~pec &           H~{\scriptsize II} &                    2 \\
                                                            IC~5179\dotfill &      22 16 09 &   $-$36 50 37 &   1.48 &   0.56 &  55 & 47.23 &  1.16 & 11.16 &    14.51 &  10.93 &             SA(rs)bc &           H~{\scriptsize II} &                  2,5 \\
                                                     ESO~420$-$G013\dotfill &      04 13 50 &   $-$32 00 25 &   0.61 &   0.58 & 110 & 48.17 &  2.28 & 11.02 &    10.27 &  10.62 &      SA0$^+$(r)~pec? &           H~{\scriptsize II} &                  4,5 \\
                                                            Arp~299\dotfill &      11 28 30 &     +58 34 10 &   1.42 &   1.25 &  28 & 48.24 &  1.05 & 11.88 &    75.60 &  11.05 &       Merger~(S~pec) &    H~{\scriptsize II} + Sy~2 &                6,8,9 \\
                                                            NGC~838\dotfill &      02 09 39 &   $-$10 08 46 &   0.62 &   0.43 &  95 & 50.78 &  2.23 & 11.00 &     9.81 &  10.53 &     SA0$^0$(rs)~pec? &           H~{\scriptsize II} &                   10 \\
                                                           NGC~5135\dotfill &      13 25 44 &   $-$29 50 01 &   1.75 &   0.86 & 125 & 52.87 &  4.35 & 11.17 &    14.72 &  10.97 &              SB(s)ab &                         Sy~2 &                  4,5 \\
                                                         NGC~5394/5\dotfill &      13 58 38 &     +37 25 28 &   1.53 &   0.80 &   5 & 54.02 &  1.05 & 11.00 &     9.90 &  11.09 &           SA(s)b~pec &                         Sy~2 &                    3 \\
                                                           NGC~5653\dotfill &      14 30 10 &     +31 12 56 &   0.76 &   0.70 &  75 & 55.53 &  1.20 & 11.06 &    11.51 &  10.87 &          (R')SA(rs)b &           H~{\scriptsize II} &                  2,5 \\
                                                           NGC~7771\dotfill &      23 51 25 &     +20 06 43 &   1.52 &   0.62 &  75 & 57.94 &  4.27 & 11.34 &    21.95 &  11.25 &               SB(s)a &           H~{\scriptsize II} &                  2,5 \\
                                                           NGC~3221\dotfill &      10 22 20$^a$ &     +21 34 11$^a$ &   1.62$^a$ &   0.34$^a$ & 167$^a$ & 59.46 &  1.86 & 11.00 &     9.92 &  11.00 &     SB(s)cd?~edge-on &           H~{\scriptsize II} &               \ldots \\
                                                     CGCG~049$-$057\dotfill &      15 13 13 &     +07 13 32 &   0.45 &   0.23 &  20 & 59.83 &  2.79 & 11.27 &    18.25 &  10.15 &               \ldots &           H~{\scriptsize II} &                2,7,5 \\
                                                             IC~860\dotfill &      13 15 03 &     +24 37 08 &   0.56 &   0.32 &  20 & 59.88 &  1.12 & 11.17 &    14.51 &  10.36 &                   S? &              {\sc LINER}/Sy2 &                2,7,5 \\
                                                             NGC~23\dotfill &      00 09 53 &     +25 55 26 &   1.13 &   0.54 & 155 & 60.52 &  3.86 & 11.05 &    11.20 &  11.08 &               SB(s)a &           H~{\scriptsize II} &                  2,5 \\
\hline
\end{tabular}
\end{center}
NOTE.---Basic properties of our complete sample of 17 LIRGs with $D<60$~Mpc and $N_{\rm H} \simlt 5 \times 10^{20}$~cm$^{-2}$, ordered by distance (given in Col.~[7]).  Col.(1): Common source name. Col.(2) and (3): Right ascension and declination, respectively. Units of right ascension are hours, minutes, and seconds, and units of declination are degrees, arcminutes, and arcseconds.  Col.(4): Semi-major axis in arcminutes. Col.(5): Semi-minor axis in arcminutes. Col.(6): Position angle in degrees, measured east from north. Values given in Col.(2)--(6) were provided by 2MASS with the exception of NGC~3221, which was given by RC3. Col.(7): Luminosity distance in Mpc, as used by Sanders \etal\ (2003). Col.(8): Galactic column density in units of $10^{20}$~cm$^{-2}$ as given by Dickey \& Lockman~(1990). Col.(9): Logarithm of the total infrared (8--1000~$\mu$m) luminosity as given by Sanders \etal\ (2003). Col.(10): Star-formation rate in units of \sfr. Col.(11): Logarithm of the stellar mass in units of $M_\odot$.  Col.(10) and (11) were computed following the methods presented in $\S$2.2. Col.(12): Optical morphological classification.  When available, these classifications are primarily based on RC3, with the exception of Arp~299, which was based on the Revised Shapley-Ames Catalog Of Bright Galaxies (Sandage \& Tammann~1981). Col.(13): Optical spectroscopic classification of source type. Col.(14): Literature reference for column~(13); 1 = Veron-Cetty \& Veron~(1986), 2 = Veilleux \etal\ (1995), 3 = Ho \etal\ (1997), 4 = Kewley \etal\ (2001), 5 = Yuan \etal\ (2010), 6 = Armus \etal\ (1989), 7 = Baan \etal\ (1998), 8 = Coziol \etal\ (1998), 9 = Garc{\'{\i}}a-Mar{\'{\i}}n \etal\ (2006) and 10 = Keel \etal\ (1985). \\
\noindent$^a$ Values taken from RC3.
}
\end{table*}

%%%%%%%%%%%%%%%%%%%%%%%%%%%%%%%%%%%%%%%%%%%%%%%%%%%%%%%%%%%%%%%%%%%%%%%%%%%%%%%%%%

In normal star-forming galaxies (i.e., those that are not dominated by luminous
active galactic nuclei [AGNs]), \xray\ emission originates from \xray\
binaries, supernovae, supernova remnants, hot ($\approx$\hbox{0.2--1~keV})
interstellar gas, and O-stars (see, e.g., Fabbiano~1989, 2006 for reviews).
Several investigations have now revealed the presence of a strong correlation
between the galaxy-wide \xray\ emission and total SFR (hereafter, the \xray/SFR
correlation; see, e.g., Bauer \etal\ 2002; Grimm \etal\ 2003; Ranalli \etal\
2003; Gilfanov \etal\ 2004a,b; Persic \etal\ 2004; Persic \& Rephaeli 2007;
Lehmer \etal\ 2008).  The \xray/SFR correlation is thought to be driven
primarily by \xray\ binary emission, and at high \xray\ energies
(\hbox{2--10~keV}), where the emission intensity from hot interstellar gas and
young stars decreases sharply, \xray\ binaries significantly dominate the
galaxy-wide \xray\ power and therefore correlate strongly with SFR (e.g.,
Persic \& Rephaeli~2002).  

Detailed \chandra\ studies of nearby star-forming late-type galaxies and
passive early-type galaxies have shown that the \xray\ point source emission
from relatively young ($\simlt$100~Myr) high-mass \xray\ binaries (HMXBs) and
older ($\simgt$1~Gyr) low-mass \xray\ binaries (LMXBs) correlate well with
galaxy SFR and stellar mass ($M_\star$), respectively (e.g., Grimm \etal\ 2003;
Colbert \etal\ 2004; Gilfanov~2004).  Colbert \etal\ (2004; hereafter C04)
utilized \chandra-resolved point source populations in 32 nearby spiral and
elliptical galaxies to measure the galaxy-wide \xray\ power due to point
sources $L_{\rm XP}$ and characterize its correlation with both SFR and
$M_\star$.  The C04 study, which focused on representative nearby galaxies with
SFR~$\approx$~0.01--10~\sfr\ and $M_\star \approx 10^{8}$--$10^{11} M_\odot$,
showed that $L_{\rm XP}$ is linearly correlated with both SFR and $M_\star$ and
the relative contributions from HMXBs and LMXBs are directly dependent on
specific-SFR (i.e., SFR/$M_\star$; a measure of the stellar-growth rate).  From
this perspective, the more commonly used \xray/SFR correlation is likely to
have significant scatter due to variations in $M_\star$ for a given SFR.  

A recent study by Persic \& Rephaeli (2007; hereafter, PR07) noted that
ultraluminous infrared galaxies (ULIRGs), which have SFR~$\simgt$~100~\sfr,
have \hbox{2--10~keV} luminosity-to-SFR ratios that are $\sim$5 times lower
than more typical nearby galaxies that have SFR~$\ll$~100~\sfr.  PR07 suggest that the
\hbox{2--10~keV} emission in ULIRGs is likely to be dominated by HMXBs with
negligible contributions from LMXBs, which likely play a more significant role
in galaxies with SFR~$\ll$~100~\sfr.  This trend has also been substantiated by
recent \chandra\ observations of powerful LIRGs and ULIRGs from the Great
Observatories All-sky LIRG Survey (GOALS; Iwasawa \etal\ 2009; hereafter,
Iw09).  Despite the improved understanding of how the galaxy-wide
\hbox{2--10~keV} emission depends on SFR and $M_\star$ from C04 and suggestions
that extremely powerful LIRGs and ULIRGs have \xray\ power dominated by young
HMXB populations (from e.g., PR07 and Iw09), researchers have yet to test
directly and constrain the C04 relation using observations of galaxies with
SFR~$\simgt 10$~\sfr.

%
%%%%%%%%%%%%%%%%%%%%%%%%%%%%%%%%%%%%%%%%%%%%%%%%%%%%%%%%%%%%%%%%%%%%%%%%%%%%%%%%%%
% Figure 1
%%%%%%%%%%%%%%%%%%%%%%%%%%%%%%%%%%%%%%%%%%%%%%%%%%%%%%%%%%%%%%%%%%%%%%%%%%%%%%%%%%
%

\begin{figure*}
\figurenum{1}
\centerline{
\includegraphics[width=19cm]{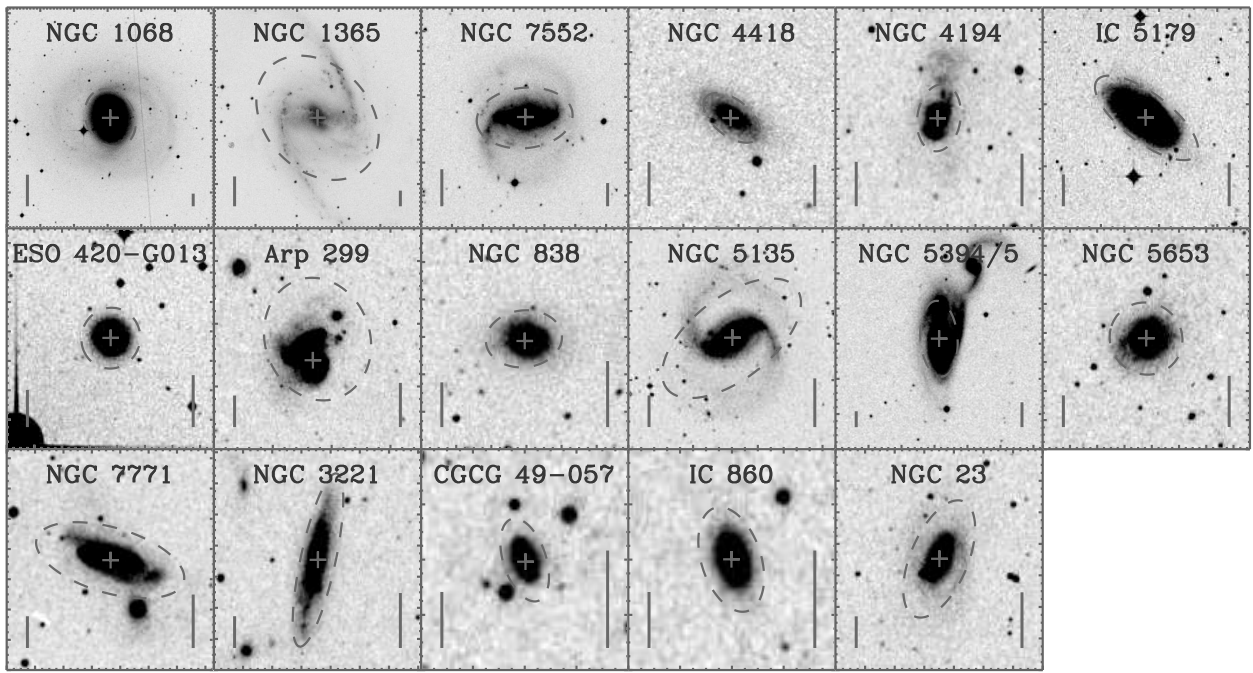}
}
\caption{
Digitized sky survey (DSS) red images of each of the 17 LIRGs that make up our
sample.  The panel sizes presented here vary to illustrate galaxy morphology
most clearly, and in the lower left and right corners of each panel, we have
provided vertical reference bars with physical sizes of 10~kpc and angular
sizes of 1~arcmin, respectively.  Dashed gray ellipses represent the total
$K_s$-band galaxy size and orientation as described in \hbox{columns~(2)--(6)}
of Table~1 and gray crosses indicate the adopted locations for the galactic
nuclei (see $\S$4.2 for details).  These images show that the galaxies making
up our LIRG sample have a diversity of optical morphological types, physical
sizes, and inclinations.  
}
\end{figure*}
%%%%%%%%%%%%%%%%%%%%%%%%%%%%%%%%%%%%%%%%%%%%%%%%%%%%%%%%%%%%%%%%%%%%%%%%%%%%%%%%%%

In this paper, we present new \chandra\ results for a volume-limited sample ($D
< 60$~Mpc) of 17 LIRGs (including 10 new LIRGs from this study; see Tables~1
and 2) with Galactic column densities \hbox{$N_{\rm H} \simlt 5 \times
10^{20}$~cm$^{-2}$} and $L_{\rm IR} \approx$1--8~$\times 10^{11} L_\odot$
(\hbox{SFR~$\simgt 10$~\sfr}; see Figs.~1 and 2).  We first utilize the
high-resolution \chandra\ data to separate spatially and measure contributions
from normal galaxy-wide \xray\ emission (i.e., not due to an AGN) and AGN when
present; this enables us both to investigate the \xray/SFR correlation in
detail for LIRGs and to characterize the AGN activity in the population.  Using
the new LIRG constraints in combination with both the \chandra\ results from
C04 for less actively star-forming normal galaxies and Iw09 for more actively
star-forming LIRGs and ULIRGs, we provide improved constraints upon how the
normal-galaxy \hbox{2--10~keV} luminosity depends on SFR and $M_\star$, and
therefore the relative contributions from HMXBs and LMXBs.  This analysis has
the advantage over previous studies in that it uniquely covers $\approx$4
orders of magnitude in SFR and $\approx$3.5 orders of magnitude in $M_\star$
and is uniformly based on \chandra\ data.

Throughout this paper, we make estimates of SFR and $M_\star$ using a
Kroupa~(2001) initial mass function (IMF); when making comparisons between
these estimates and those quoted in other studies, we have adjusted all values
to correspond to our adopted IMF.  We sometimes make comparisons with the Milky
Way (MW).  We adopt values of $M_\star \approx 5 \times 10^{10}$~$M_\odot$
(Hammer \etal\ 2007), SFR~$\approx 2.0$~\sfr\ (McKee \& Williams~1997), and
\hbox{2--10~keV} luminosity $L_{\rm HX}^{\rm gal} = 3.3 \times 10^{39}$~\xlum\
(Grimm \etal\ 2002) for the MW $H_0$ = 70~\hbox{km s$^{-1}$ Mpc$^{-1}$},
$\Omega_{\rm M}$ = 0.3, and $\Omega_{\Lambda}$ = 0.7 are adopted throughout
this paper (e.g., Spergel \etal\ 2003).

%
%%%%%%%%%%%%%%%%%%%%%%%%%%%%%%%%%%%%%%%%%%%%%%%%%%%%%%%%%%%%%%%%%%%%%%%%
\section{Sample}
%%%%%%%%%%%%%%%%%%%%%%%%%%%%%%%%%%%%%%%%%%%%%%%%%%%%%%%%%%%%%%%%%%%%%%%%
%

\subsection{LIRG Sample Selection}

Using the IRAS Revised Bright Galaxies Sample (RBGS; Sanders \etal\ 2003), we
constructed a sample of nearby LIRGs that were at distances less than 60~Mpc
and had Galactic column densities \hbox{$N_{\rm H} < 5 \times
10^{20}$~cm$^{-2}$} (Dickey \& Lockman~1990).  The former requirement results
in {\it Chandra} imaging resolution of $\simlt$150~pc in the region of the
nucleus, which allows for reasonable distinction between nuclear AGNs and
starburst regions (see, e.g., Levenson \etal\ 2004 for an analysis of the AGN
plus starburst nucleus of NGC~5135 in our sample).  The latter requirement results in
moderate-to-good Galactic transparency to soft \xray\ emission, which is
valuable for detecting and modeling the emission from hot
\hbox{$\approx$0.5--1~keV} gas that is often found in star-forming galaxies.
In total, 36 RBGS LIRGs reside at $D < 60$~Mpc and 17 satisfy our Galactic
column density criterion.  In Table~1, we provide the basic properties of our
LIRG sample, and in Figure~1, we show Digitized Sky Survey (DSS) red images of
the galaxies in our sample.  In contrast to ULIRGs, which are nearly all
late-type galaxies with evidence for mergers/interactions (e.g., Melnick \&
Mirabel~1990; Sanders \etal\ 1988; Chen \etal\ 2010), the LIRGs that make up
our sample consist of a variety of optical morphological types (e.g., spirals,
barred spirals, S0s) with only a small fraction ($\sim$10\%) showing evidence
for mergers/interactions.

In Figure~2, we show the infrared luminosity $L_{\rm IR}$ versus distance for
the RBGS and highlight our sample ({\it filled black circles\/}).  Of the 17
LIRGs in our sample, seven (NGC~1068, NGC~1365, NGC~7552, NGC~4418, NGC~4194,
Arp~299, and NGC~5135) had \chandra\ archival data.  Detailed \chandra\
analyses for these sources can be found in the literature for NGC~1068 (Young
\etal\ 2001; Smith \& Wilson~2001), NGC~1365 (Risaliti \etal\ 2009a,b; Soria
\etal\ 2009; Strateva \& Komossa~2009; Wang \etal\ 2009), NGC~4194 (Kaaret \&
Alonso-Herrero~2008), Arp~299 (Zezas \etal\ 2003), and NGC~5135 (Levenson
\etal\ 2004).  To enable new \chandra\ constraints on the LIRG population in
general, we proposed successfully in \chandra\ Cycle~10 (PI: D.~M.~Alexander)
to observe the remaining 10 LIRGs.  Our \chandra\ program (and consequently
adopted exposure times; see Table~2) was designed to allow for the detection of
ultraluminous \xray\ sources (ULXs) down to a \hbox{0.5--8~keV} limit of
$\approx$$2 \times 10^{39}$~\xlum\ for the 17 LIRGs.  In a forthcoming paper
(Jackson \etal\ in-prep), we will present the \xray\ and statistical properties
of the ULX populations detected in the LIRG sample.

As is evident from Figure~2, our sample provides a useful complement to the
$L_{\rm IR}$ distribution of \chandra-observed galaxies between less
active nearby galaxies (e.g., from C04) and more powerful LIRGs and ULIRGs
(e.g., from the GOALS; Iw09).  We note that the RBGS sources with \chandra\
observations highlighted in Figure~2 do {\it not} constitute a complete census
of the \chandra\ archive.  However, incompleteness is only most severe for
galaxies with $L_{\rm IR} \simlt 10^{11} L_\odot$, which are not the primary
focus of this analysis; detailed \chandra\ and multiwavelength studies of such
galaxies will be the subject of forthcoming work by the \spitzer\ Infrared
Nearby Galaxies Survey (SINGS) collaboration (Jenkins \etal\ in-prep; Ptak
\etal\ in-prep; see Kennicutt \etal\ 2003 for survey details).  For LIRGs and
ULIRGs, the \chandra\ observations highlighted in Figure~2 are nearly complete. 

%
%%%%%%%%%%%%%%%%%%%%%%%%%%%%%%%%%%%%%%%%%%%%%%%%%%%%%%%%%%%%%%%%%%%%%%%%%%%%%%%%%%
% Figure 2
%%%%%%%%%%%%%%%%%%%%%%%%%%%%%%%%%%%%%%%%%%%%%%%%%%%%%%%%%%%%%%%%%%%%%%%%%%%%%%%%%%
%

\begin{figure}
\figurenum{2}
\centerline{
\includegraphics[width=9cm]{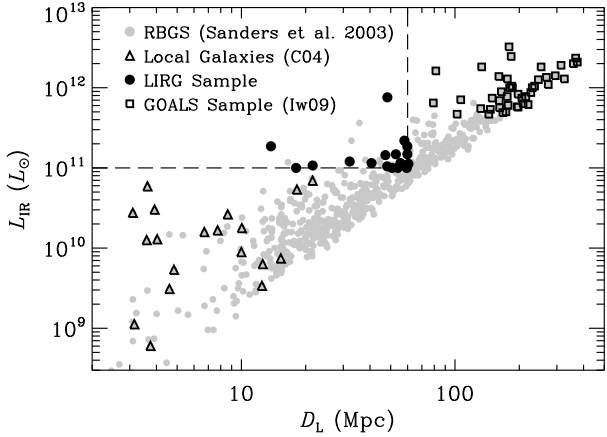}
}
\caption{
Total infrared (8--1000$\mu$m) luminosity $L_{\rm IR}$ versus distance for IRAS
sources ({\it light-gray filled circles\/}), as provided by Sanders \etal\
(2003).  Our complete sample of \chandra-observed LIRGs that have $D < 60$~Mpc
and \hbox{$N_{\rm H} \simlt 5 \times 10^{10}$~cm$^{-2}$} ({\it dashed boundary
lines\/}) are indicated with black filled circles.  Nearby normal late-type
galaxies that have been studied extensively with \chandra\ have been
highlighted with open triangles (C04) and LIRGs/ULIRGs from the Great
Observatories All-sky LIRG Survey (GOALS) sample (Iw09) are indicated with open
squares. Our sample provides new constraints in the range between
nearby normal galaxies and powerful LIRGs/ULIRGs.
}
\end{figure}
%%%%%%%%%%%%%%%%%%%%%%%%%%%%%%%%%%%%%%%%%%%%%%%%%%%%%%%%%%%%%%%%%%%%%%%%%%%%%%%%%%

\subsection{Stellar Mass and Star Formation Rate Estimates}

We estimated stellar masses for the LIRGs in our sample following the
prescription outlined in Appendix~2 of Bell \etal\ (2003), which provides
galaxy mass-to-light ratios in several optical/near-IR bands as a function of
optical/near-IR color.  We utilized $B-V$ colors from the NASA/IPAC
Extragalactic Database (NED), which were primarily from the Third Reference
Catalog of Bright Galaxies (RC3; de Vaucouleurs \etal\ 1991).  $K$-band
luminosities were computed using data from the Two Micron All Sky Survey
(2MASS) large galaxy atlas (Jarrett \etal\ 2003) and $K_s$-band extended source
catalogs.\footnote{The 2MASS extended source catalog is available online via
http://tdc-www.harvard.edu/catalogs/tmxsc.html.}  We note that for NGC~3221, no
2MASS counterpart was available, and we therefore adopted the $M_\star$
estimate provided by the GOALS team (Howell \etal\ 2010; hereafter, H10),
adjusted to match our adopted distance and IMF (see below).  For the 16 LIRGs
having 2MASS counterparts, we computed stellar masses using the following
equation:
\begin{equation}
\log M_\star/M_\odot = \log L_K/L_{K,\odot} + 0.135(B-V) - 0.356.
\end{equation}
The numerical constants in equation~1 were supplied by Table~7 of Bell et al.
(2003) and are appropriate for our choice of $B-V$ color and $L_K$; the
normalization has been adjusted by $−$0.15 dex to account for our adopted
Kroupa (2001) IMF.  Uncertainties at the $\approx$0.1~dex level are expected due
to variations in stellar ages, dust attenuation, and bursts of star formation
(see Bell \etal\ 2003 for further details).

%%%%%%%%%%%%%%%%%%%%%%%%%%%%%%%%%%%%%%%%%%%%%%%%%%%%%%%%%%%%%%%%%%%%%%%%%%%%%%%%%%
% Table 2
%%%%%%%%%%%%%%%%%%%%%%%%%%%%%%%%%%%%%%%%%%%%%%%%%%%%%%%%%%%%%%%%%%%%%%%%%%%%%%%%%%
\begin{table*}
\begin{center}
\caption{\chandra\ Observation Log}
\begin{tabular}{lccccccc}
\hline\hline
  &  & {\sc Obs. Start} & {\sc Exposure Time}$^a$ & & & {\sc Subarray?}$^c$  &  \\
\multicolumn{1}{c}{\sc Source Name} & {\sc Obs. ID} & (UT) & (ks) & {\sc Obs. Mode}$^b$ & {\sc Camera} & (Y/N) & {\sc Pipeline Version}$^d$ \\
\hline\hline
                   NGC~1068\ldots\ldots\ldots\ldots\ldots\ldots\ldots\ldots &          344 &        2000 Feb 21, 15:48 &   46.9 &         F &    ACIS-S &         N &        7.6.11.2 \\
                                                                            &          370 &        2000 Feb 22, 05:51 &    3.7 &        VF &    ACIS-S &         Y &        7.6.11.1 \\
                                               NGC~1365\ldots\ldots\dotfill &         3554 &        2002 Dec 24, 14:57 &   13.6 &         F &    ACIS-S &         N &           7.6.9 \\
                                                                            &     6868$^e$ &        2006 Apr 17, 19:05 &   14.6 &         F &    ACIS-S &         Y &         7.6.7.1 \\
                                                                            &     6869$^e$ &        2006 Apr 20, 10:00 &   15.5 &         F &    ACIS-S &         Y &         7.6.7.1 \\
                                                                            &     6870$^e$ &        2006 Apr 23, 10:05 &   14.4 &         F &    ACIS-S &         Y &         7.6.7.2 \\
                                                                            &     6871$^e$ &        2006 Apr 10, 07:02 &   13.2 &         F &    ACIS-S &         Y &         7.6.7.1 \\
                                                                            &     6872$^e$ &        2006 Apr 12, 12:31 &   14.6 &         F &    ACIS-S &         Y &         7.6.7.1 \\
                                                                            &     6873$^e$ &        2006 Apr 14, 23:26 &   14.6 &         F &    ACIS-S &         Y &         7.6.7.1 \\
                                               NGC~7552\ldots\ldots\dotfill &         7848 &        2007 Mar 31, 05:48 &    5.1 &         F &    ACIS-S &         N &          7.6.11 \\
                                               NGC~4418\ldots\ldots\dotfill &         4060 &        2003 Mar 10, 00:58 &   19.8 &         F &    ACIS-S &         Y &           7.6.8 \\
                                                                            &        10391 &        2009 Feb 20, 20:34 &    5.7 &        VF &    ACIS-S &         Y &       7.6.11.10 \\
                                               NGC~4194\ldots\ldots\dotfill &         7071 &        2006 Sep 09, 23:30 &   35.0 &         F &    ACIS-S &         N &         7.6.8.1 \\
                                                IC~5179\ldots\ldots\dotfill &        10392 &        2009 Jun 21, 13:51 &   12.0 &        VF &    ACIS-S &         Y &       7.6.11.10 \\
                                         ESO~420$-$G013\ldots\ldots\dotfill &        10393 &        2009 May 13, 05:46 &   12.4 &        VF &    ACIS-S &         Y &       7.6.11.10 \\
                                                Arp~299\ldots\ldots\dotfill &         1641 &        2001 Jul 13, 11:07 &   24.3 &         F &    ACIS-I &         Y &           7.6.9 \\
                                                NGC~838\ldots\ldots\dotfill &        10394 &        2008 Nov 23, 09:03 &   13.8 &        VF &    ACIS-S &         Y &        7.6.11.9 \\
                                               NGC~5135\ldots\ldots\dotfill &         2187 &        2001 Sep 04, 15:23 &   26.4 &         F &    ACIS-S &         Y &           7.6.9 \\
                                             NGC~5394/5\ldots\ldots\dotfill &        10395 &        2009 Mar 18, 23:28 &   15.7 &        VF &    ACIS-S &         Y &       7.6.11.10 \\
                                               NGC~5653\ldots\ldots\dotfill &        10396 &        2009 Apr 11, 05:46 &   16.5 &        VF &    ACIS-S &         Y &       7.6.11.10 \\
                                               NGC~7771\ldots\ldots\dotfill &        10397 &        2009 May 22, 05:12 &   16.7 &        VF &    ACIS-S &         Y &       7.6.11.10 \\
                                               NGC~3221\ldots\ldots\dotfill &        10398 &        2009 Mar 19, 04:19 &   19.0 &        VF &    ACIS-S &         Y &       7.6.11.10 \\
                                         CGCG~049$-$057\ldots\ldots\dotfill &        10399 &        2009 Apr 17, 05:43 &   19.1 &        VF &    ACIS-S &         Y &       7.6.11.10 \\
                                                 IC~860\ldots\ldots\dotfill &        10400 &        2009 Mar 24, 02:38 &   19.1 &        VF &    ACIS-S &         Y &       7.6.11.10 \\
                                                 NGC~23\ldots\ldots\dotfill &        10401 &        2008 Oct 27, 06:22 &   19.4 &        VF &    ACIS-S &         Y &        7.6.11.9 \\
\hline
\end{tabular}
\end{center}
NOTE.---Links to the data sets in this table have been provided in the electronic edition. \\
$^a$ All observations were continuous. These times have been corrected for removed data that was affected by high background; see $\S$~3.\\
$^b$ The observing mode (F=Faint mode; VF=Very Faint mode).\\
$^c$ States whether a subarray was used in the observation.\\
$^d$ The version of the CXC pipeline software used for basic processing of the data.\\
$^e$ Observation covers only a small fraction of the galactic extent and was therefore not used in the analyses presented in this paper.
\end{table*}

We note that 12 of the LIRGs in our sample (including 11 LIRGs with stellar
mass estimates from equation~1) have been studied extensively from the far-UV
to IR (including observations with both \galex\ and \spitzer) by the GOALS
collaboration and have stellar mass estimates available (e.g., Armus \etal\
2009; H10).  After accounting for known differences in adopted distances and
IMF, we find that our stellar mass estimates are in good agreement with those
of H10, with our values being larger by $\approx$4\% (median value) with a
1$\sigma$ scatter of $\approx$0.09~dex.

By selection, our LIRG sample consists of galaxies radiating powerfully over
the far-IR (\hbox{8--1000}~$\mu$m) wavelength range.  The far-IR light is
expected to be produced by dust-reprocessed emission from underlying
UV-obscured star-formation activity and can therefore be used as a direct
measurement of the SFR.  Using the \hbox{8--1000}~$\mu$m total infrared
luminosities $L_{\rm IR}$ provided by Sanders \etal\ (2003) and the methods
described in $\S$~3.2 of Bell \etal\ (2005), we estimated the total galaxy SFR
using the following equation:
\begin{equation}
{\rm SFR}(M_\odot~{\rm yr}^{-1}) = \gamma 9.8 \times 10^{-11} L_{\rm IR},
\end{equation}
where $L_{\rm IR}$ is expressed in units of solar bolometric luminosity
($L_\odot = 3.9 \times 10^{33}$~ergs~s$^{-1}$) and the factor \hbox{$\gamma
=$~1.000--1.024} provides small corrections that account for UV emission
emerging from unobscured star-forming regions.  Equation~2 is a direct variant
of equation~1 presented by Bell \etal\ (2005) and was derived from {\ttfamily
P\'{E}GASE} stellar-population models, which assumed a 100~Myr old population
with constant SFR and a Kroupa~(2001) IMF.  This calibration uses identical
assumptions to the SFR calibrations provided by Kennicutt~(1998).  For the 12
LIRGs that overlap with the H10 sample, $\gamma$ was provided by H10 using
values of the total infrared flux from IRAS ($f_{\rm IR}$) and the far-UV flux
($f_{\rm FUV}$) from \galex: $\gamma \equiv 1 + f_{\rm FUV}/f_{\rm IR}$.  For
these 12 LIRGs, we find good agreement between SFR measurements, with our SFR
values being larger by $\approx$4\% (median offset) with small scatter
($\approx$0.003~dex; 1$\sigma$).  For the five LIRGs in our sample without H10
counterparts, we adopted $\gamma^{\rm median} \approx 1.017$, the median value
of $\gamma$ for the 12 LIRGs with H10 counterparts.

In Table~1, we provide the derived stellar masses and SFRs for our LIRGs.  The
sample spans $M_\star \approx$ \hbox{(1.3--16)}~$\times 10^{10} M_\odot$
(median value of $M_\star \approx 7.2 \times 10^{10} M_\odot$) and
SFR~$=$~9.8--75.6~\sfr\ (median value of SFR~$\approx 11$~\sfr).  For
comparison, the LIRGs/ULIRGs from the \chandra\ GOALS project (Iw09) span a
similar range of stellar masses ($M_\star \approx$ \hbox{(1.6--27)}~$\times
10^{10} M_\odot$; median value of $M_\star$~$\approx 6.5 \times 10^{10}
M_\odot$), but contain more actively star-forming galaxies (SFR~$\approx$
\hbox{14--215}~\sfr; median value of SFR~$\approx 68$~\sfr).  The nearby
late-type galaxies in the C04 study span the stellar mass range of $\approx$$6
\times 10^7$--$10^{11} M_\odot$ (median value of $M_\star$~$\approx 1.8 \times
10^{10} M_\odot$) and less powerful star-forming galaxies with
SFR~=~0.03--5.6~\sfr\ (median value of SFR = 0.8~\sfr).  

%%%%%%%%%%%%%%%%%%%%%%%%%%%%%%%%%%%%%%%%%%%%%%%%%%%%%%%%%%%%%%%%%%%%%%%%%%%%%%%%%%
% Table 3
%%%%%%%%%%%%%%%%%%%%%%%%%%%%%%%%%%%%%%%%%%%%%%%%%%%%%%%%%%%%%%%%%%%%%%%%%%%%%%%%%%
\begin{table*}
\begin{center}
\caption{Basic X-ray Properties and Spectral Fitting Results}
\begin{tabular}{lcccccccccc}
\hline\hline
 &  \multicolumn{2}{c}{\sc Net Counts} &  &  & & & & & Other &  AGN  \\
 &  \multicolumn{2}{c}{\rule{1in}{0.01in}}  &  & $kT$ & &  & $\log L_{\rm HX}^{\rm gal}$ & & Evidence & Dominant? \\
\multicolumn{1}{c}{\sc Source Name} &  (0.5--8~keV) & (2--8~keV) & $\Gamma$ & (keV) & $\chi^2/\nu$ & $\nu$ & (\xlum) &  $\Phi_{\rm 2-8~keV}^{\rm nuc}/\Phi_{\rm 2-8~keV}^{\rm gal}$ & for AGN & (Y/N) \\
 \multicolumn{1}{c}{(1)} &  (2) & (3) & (4)  & (5) & (6) & (7) & (8) & (9) & (10) & (11) \\
\hline\hline
                   NGC~1068\ldots\ldots\ldots &               150875 $\pm$ 404 &                11215 $\pm$ 131 &                         \ldots &                         \ldots &                         \ldots &                         \ldots &                                     41.49$^a$ &                                      0.74$^a$ &                               Opt., IR, X-ray &                                             Y \\
                             NGC~1365\dotfill &                 4719 $\pm$ 127 &                   841 $\pm$ 90 &                1.88 $\pm$ 0.12 &                0.73 $\pm$ 0.05 &                           1.26 &                             84 &                                         40.65 &                                          0.88 &                               Opt., IR, X-ray &                                             Y \\
                             NGC~7552\dotfill &                   939 $\pm$ 40 &                   113 $\pm$ 23 &                2.06 $\pm$ 0.13 &                0.64 $\pm$ 0.04 &                           0.91 &                             46 &                                         40.40 &                                       $<$0.08 &                                        \ldots &                                             N \\
                             NGC~4418\dotfill &                    49 $\pm$ 23 &                          $<$53 &                3.28 $\pm$ 0.46 &                         \ldots &                         \ldots &                         \ldots &                                      $<$39.58 &                                        \ldots &                                        \ldots &                                             N \\
                             NGC~4194\dotfill &                  2413 $\pm$ 53 &                   263 $\pm$ 24 &                2.11 $\pm$ 0.08 &                0.83 $\pm$ 0.02 &                           1.23 &                             79 &                                         40.43 &                                          0.50 &                                            IR &                                             Y \\
                              IC~5179\dotfill &                   520 $\pm$ 33 &                    62 $\pm$ 22 &                2.10 $\pm$ 0.19 &                0.84 $\pm$ 0.06 &                           0.62 &                             32 &                                         40.49 &                                       $<$0.09 &                                        \ldots &                                             N \\
                       ESO~420$-$G013\dotfill &                   762 $\pm$ 31 &                    53 $\pm$ 15 &                3.00 $\pm$ 0.19 &                0.71 $\pm$ 1.61 &                           1.09 &                             30 &                                         40.08 &                                          0.43 &                                      Opt., IR &                                             N \\
                              Arp~299\dotfill &                  4456 $\pm$ 72 &                   836 $\pm$ 36 &                1.85 $\pm$ 0.06 &                0.80 $\pm$ 0.02 &                           1.29 &                            130 &                                         41.32 &                                          0.11 &                               Opt., IR, X-ray &                                             N \\
                              NGC~838\dotfill &                   632 $\pm$ 29 &                    70 $\pm$ 14 &                2.07 $\pm$ 0.13 &                0.73 $\pm$ 2.84 &                           0.83 &                             28 &                                         40.61 &                                          0.41 &                                        \ldots &                                             N \\
                             NGC~5135\dotfill &                  3756 $\pm$ 74 &                   346 $\pm$ 37 &                2.53 $\pm$ 0.08 &                0.78 $\pm$ 0.55 &                           1.35 &                            117 &                                         40.70 &                                          0.44 &                               Opt., IR, X-ray &                                             N \\
                           NGC~5394/5\dotfill &                   289 $\pm$ 38 &                    92 $\pm$ 30 &                1.69 $\pm$ 0.38 &                0.09 $\pm$ 0.20 &                           0.65 &                             37 &                                         40.49 &                                          0.27 &                                          Opt. &                                             N \\
                             NGC~5653\dotfill &                   389 $\pm$ 30 &                    50 $\pm$ 19 &                2.21 $\pm$ 0.26 &                0.68 $\pm$ 0.06 &                           1.20 &                             26 &                                         40.18 &                                       $<$0.11 &                                        \ldots &                                             N \\
                             NGC~7771\dotfill &                   864 $\pm$ 42 &                   233 $\pm$ 30 &                1.24 $\pm$ 0.21 &                0.63 $\pm$ 2.39 &                           1.03 &                             55 &                                         41.18 &                                          0.53 &                                     IR, X-ray &                                             Y \\
                             NGC~3221\dotfill &                   310 $\pm$ 31 &                    69 $\pm$ 22 &                1.58 $\pm$ 0.16 &               $\approx$0.0$^c$ &                           0.95 &                             26 &                                         40.66 &                                          0.40 &                                        \ldots &                                             N \\
                       CGCG~049$-$057\dotfill &                    33 $\pm$ 13 &                          $<$29 &                1.05 $\pm$ 0.57 &                         \ldots &                         \ldots &                         \ldots &                                      $<$40.29 &                                        \ldots &                                        \ldots &                                             N \\
                               IC~860\dotfill &                          $<$44 &                          $<$35 &               $\approx$2.0$^b$ &                         \ldots &                         \ldots &                         \ldots &                                      $<$40.19 &                                        \ldots &                                        \ldots &                                             N \\
                               NGC~23\dotfill &                   717 $\pm$ 38 &                          $<$71 &                2.03 $\pm$ 0.21 &                0.73 $\pm$ 1.12 &                           1.18 &                             41 &                                      $<$40.54 &                                       $>$0.24 &                                        \ldots &                                             N \\
\hline
\end{tabular}
\end{center}
NOTE.---Basic X-ray properties and spectral fitting results for the LIRGs in our sample.  Col.(1): Source name. Col.(2) and (3): Net source counts in the \hbox{0.5--8~keV} and \hbox{2--8~keV} bandpasses, respectively (see $\S$4.1 for details).  For sources with $\simgt$300 FB counts listed in column~2, \xray\ spectral fitting was performed over the \hbox{0.5--8~keV} band. Col.(4) and (5): Best-fit values of $\Gamma$ and $kT$, respectively, for an assumed power-law plus Raymond-Smith thermal plasma model (see $\S$4.1 for details). Col.(6) and (7): Reduced $\chi^2$ value ($\chi^2$/$\nu$) and the number of degrees of freedom in the fits ($\nu$), respectively.  Col.(8): Integrated \hbox{2--10~keV} total luminosity, computed using our best-fit models and distances provided in Table~1, column~(7). Col.(9): Ratio of \hbox{2--8~keV} count-rates from the nuclear region of each galaxy and the total galaxy-wide extent.  Count-rates were computed following the methods described in $\S\S$4.1 and 4.2. A source with large nuclear contributions to the total \hbox{2--8~keV} count-rates harbors either an AGN or a strong nuclear starburst. Col.(10): Notes whether there is evidence for AGN activity via optical spectroscopy (``Opt''; see columns~13 and 14 in Table~1), the mid-infrared spectrum (``IR''), or from \xray\ studies in the literature (``X-ray''); see $\S$~4.2.  Col.(11): Indicates whether the integrated \hbox{2--10~keV} luminosity (column~8) is dominated by AGN activity.  We consider sources that have both $\Phi_{\rm 2-8~keV}^{\rm nuc}/\Phi_{\rm 2-8~keV}^{\rm gal} \ge 0.5$ (column~9) and additional evidence for AGN activity from column~10 to have the \hbox{2--10~keV} luminosity dominated by an AGN (``Y''); otherwise, we consider the galaxy to be dominated by \xray\ binary emission (``N'').\\
\noindent $^a$Values are inferred using detailed analysis by Young \etal\ (2001).\\
\noindent $^b$A simple power-law model with $\Gamma \approx 2.0$ was assumed for IC~860.\\
\noindent $^c$Best-fit spectrum did not require the use of a hot gas component.\\
\end{table*}

%%%%%%%%%%%%%%%%%%%%%%%%%%%%%%%%%%%%%%%%%%%%%%%%%%%%%%%%%%%%%%%%%%%%%%%%%%%%%%%%%%

%
%%%%%%%%%%%%%%%%%%%%%%%%%%%%%%%%%%%%%%%%%%%%%%%%%%%%%%%%%%%%%%%%%%%%%%%%
\section{{\itshape Chandra} Observations and Data Reduction}
%%%%%%%%%%%%%%%%%%%%%%%%%%%%%%%%%%%%%%%%%%%%%%%%%%%%%%%%%%%%%%%%%%%%%%%%
%

In Table~2, we present the \chandra\ ACIS observation log for our LIRG sample.
Our analyses began with the \chandra\ \xray\ Center (CXC) processed Level~2
events files, which were all processed using pipeline version 7.6.8 or greater.
Additional reductions and analysis of the data were performed using the
\chandra\ Interactive Analysis of Observations ({\ttfamily CIAO}) Version~4.1
tools and custom software.  We screened our events files for undesirable grades
using the standard \asca\ grade set (\asca\ grades 0, 2, 3, 4, 6) and excluded
from the events lists background flaring events that were $\simgt$3$\sigma$
times higher than normal.  For the observations 344 (NGC~1068), 3554
(NGC~1365), 7071 (NGC~4194) and 2187 (NGC~5135) we removed short flaring events
lasting $\approx$0.5, 1.0, 0.5, and 3.0~ks, respectively.

Out of the 17 LIRGs in our sample, three galaxies (NGC~1068, 1365, and 4418)
have more than one \chandra\ observation available.  However, in the case of
NGC~1365, observations 6868, 6869, 6870, 6871, 6872, and 6873 cover only a
small fraction of the entire galactic extent as defined in Table~1 and
constitute $\approx$86.5\% of the total ACIS exposure available.  In order to
avoid galaxy-wide spectral fits (see $\S$~4.1 below) that are statistically
biased towards these smaller non-representative regions, we therefore chose for
this study to make use of only observation 3554, which covers the entire
galactic extent and is sufficiently sensitive for our study.

For the remaining two LIRGs with more than one \chandra\ observation (NGC~1068
and 4418), we corrected the relative astrometry to match the frames with the
longest exposures.  We first ran {\ttfamily wavdetect} at a false-positive
probability threshold of $10^{-6}$ to create point-source catalogs for each
frame.  For a given galaxy, we filtered each of our {\ttfamily wavdetect}
catalogs to include only sources that were common to all frames and within
6\arcmin\ of the \chandra\ aim-point, where the point-spread function size and
resulting positional errors are small.  Using these source lists, we registered
each aspect solution and events list to the frame with the longest exposure
time using the {\ttfamily CIAO} tools {\ttfamily reproject\_aspect} and
{\ttfamily reproject\_events}, respectively.  The resulting astrometric
reprojections gave nearly negligible linear translations ($<$0.38~pixels),
rotations ($<$0.1~deg), and stretches ($<$0.06\% of the pixel size).  Using the
astrometrically-reprojected events lists, we combined the observations using
{\ttfamily dmmerge} to create merged events lists.

Using the events lists, we constructed images of each LIRG in three bands:
\hbox{0.5--8~keV} (full band; FB), \hbox{0.5--2~keV} (soft band; SB), and
\hbox{2--8~keV} (hard band; HB).  For each of the three bands, we constructed
corresponding exposure maps following the basic procedure outlined in $\S$3.2
of Hornschemeier \etal\ (2001); these maps were normalized to the effective
exposures of sources located at the aim points.  This procedure takes into
account the effects of vignetting, gaps between the CCDs, bad column and pixel
filtering, the spatially-dependent degredation of the ACIS optical blocking
filter, and the reduced area of frames that use subarrays (see Table~2).  A
photon index of $\Gamma=1.4$, the slope of the extragalactic cosmic
\hbox{X-ray} background in the full band (e.g., Hickox \& Markevitch~2006), was
assumed in creating the exposure maps.  

%
%%%%%%%%%%%%%%%%%%%%%%%%%%%%%%%%%%%%%%%%%%%%%%%%%%%%%%%%%%%%%%%%%%%%%%%%%%%%%%%%%%
% Figure 3
%%%%%%%%%%%%%%%%%%%%%%%%%%%%%%%%%%%%%%%%%%%%%%%%%%%%%%%%%%%%%%%%%%%%%%%%%%%%%%%%%%
%

\begin{figure*}
\figurenum{3}
\centerline{
\includegraphics[width=19cm]{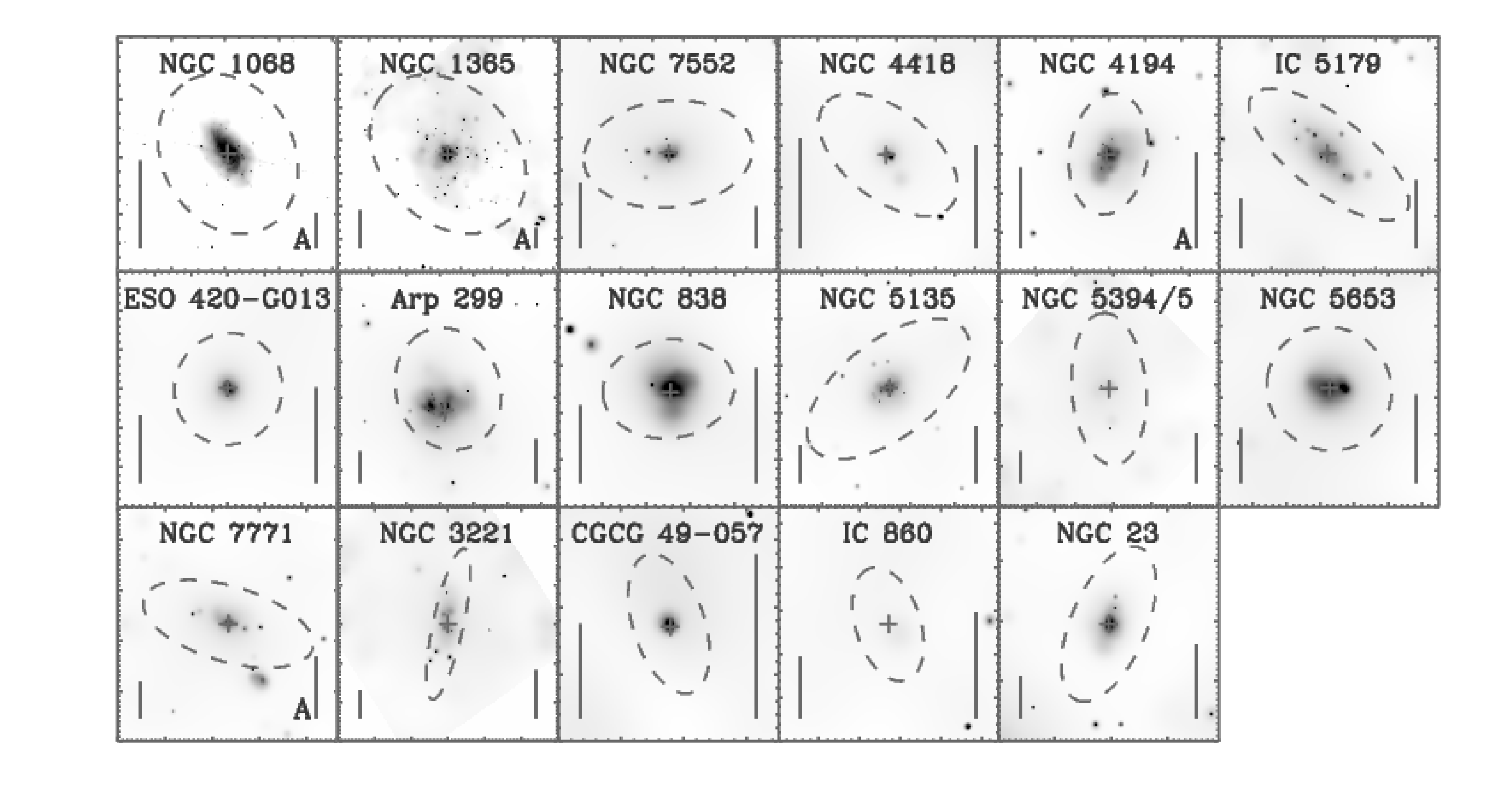}
}
\caption{
Full band (0.5--8~keV) adaptively smoothed images of the 17 LIRGs in our
sample.  The images were created using the {\ttfamily CIAO} tool {\ttfamily
csmooth}, applied to the raw images at the 2.5$\sigma$ level.  Symbols and
scales have the same meaning as they did in Figure~1; however, the image sizes
have been adjusted to show most clearly the \xray\ morphologies of the
galaxies.  Galaxies with \hbox{2--8~keV} emission dominated by an AGN
component (see $\S$~4.2) have been denoted with an ``A'' in the lower right.
We note that each image has a different exposure time (see Table~2) and the
smoothed-image sharpness generally increases with increasing exposure time.
}
\end{figure*}
%%%%%%%%%%%%%%%%%%%%%%%%%%%%%%%%%%%%%%%%%%%%%%%%%%%%%%%%%%%%%%%%%%%%%%%%%%%%%%%%%%

%
%%%%%%%%%%%%%%%%%%%%%%%%%%%%%%%%%%%%%%%%%%%%%%%%%%%%%%%%%%%%%%%%%%%%%%%%
\section{Analyses}
%%%%%%%%%%%%%%%%%%%%%%%%%%%%%%%%%%%%%%%%%%%%%%%%%%%%%%%%%%%%%%%%%%%%%%%%
%

\subsection{Extraction of Total LIRG X-ray Properties}

To measure the total galaxy-wide \xray\ emission originating from each of the
LIRGs, we began by extracting {\it total galactic counts} $S_{\rm tot}^{\rm
gal}$ and effective exposure times $T_{\rm tot}^{\rm gal}$ from each of the
three images and exposure maps using elliptical regions that approximate the
galactic stellar extent.  Whenever possible, we adopted for these elliptical
regions the ``total'' semi-major/minor axes and position angles provided by the
Two Micron All Sky Survey (2MASS) large galaxy atlas (Jarrett \etal\ 2003) and
the $K_s$-band extended source catalogs; these regions, which were determined using a
curve-of-growth analysis (see $\S$4.2 of Jarrett \etal\ 2003), are shown in
Figure~1 ({\it dashed ellipses\/}) and their values are tabulated in Table~1.
For NGC~3221, 2MASS values were not available and we therefore adopted the RC3
values for the semi-major/minor axes and position angle for this galaxy.  We
note that for 14 of the 17 LIRGs in our sample, both 2MASS and RC3 galaxy
extent parameters were available.  We find that the 2MASS values of major and
minor axes are on average larger than those of RC3 by only small factors of
$\approx$1.13 and $\approx$1.08, respectively.

To estimate the background contributions to the total counts for each galaxy,
we extracted {\it background counts} $S_{\rm bkg}^{\rm gal}$ and exposure times
$T_{\rm bkg}^{\rm gal}$ for each of the three images using \hbox{3--5} moderate
size (\hbox{$\approx$20--90~arcsec} radius) circular regions that were outside
the galactic $K_s$-band ellipses and did not overlap within 10\arcsec\ of the
galaxy or any sources detected by {\ttfamily wavdetect} at a false-positive
probability threshold of 10$^{-6}$.  

Using the galaxy and background counts, we estimated the net galaxy-wide LIRG
counts \hbox{$N^{\rm gal} = S_{\rm tot}^{\rm gal} - S_{\rm bkg}^{\rm gal}
T_{\rm tot}^{\rm gal}/T_{\rm bkg}^{\rm gal}$} for each of the three bands.
Since $T_{\rm tot}^{\rm gal}$ and $T_{\rm bkg}^{\rm gal}$ are computed using
the sum of exposure map values over the source and background regions,
respectively, the estimated background contribution to the net counts (i.e.,
$S_{\rm bkg}^{\rm gal} T_{\rm tot}^{\rm gal}/T_{\rm bkg}^{\rm gal}$) are
corrected for both extraction area and vignetting.  We note that while these
background estimates will account for emission due to unrelated background
\xray\ sources (e.g., distant AGNs; see, e.g., Morretti \etal\ 2003; Bauer
\etal\ 2004; Kim \etal\ 2007) that fall below our source detection threshold,
these estimates do {\it not} include contributions from unrelated sources that
are detected in the \xray\ band that may lie within the extent of the galaxy.
Using the best-fit $\log N$--$\log S$ from Morretti \etal\ (2003) and the
elliptical galaxy areas described in columns~(2)--(6) of Table~1, we estimate
that unrelated \xray\ detected sources are expected to contribute $\simlt$8\%
(median $\approx$2\%) to the total \hbox{2--8~keV} band count-rate of each LIRG
in our sample.  Since these contributions are small and subject to large
uncertainties, we therefore do not make any additional corrections to the net
galaxy count-rates.

We considered a source to be {\it detected} in a given band if the
signal-to-noise ratio (\hbox{S/N = $N^{\rm gal}/[S_{\rm bkg}^{\rm gal} T_{\rm
tot}^{\rm gal}/T_{\rm bkg}^{\rm gal}]^{0.5}$}) was greater than or equal to
three.  When a source was not detected in a given band, we computed 3$\sigma$
upper limits on its net counts.  In columns~(2) and (3) of Table~3, we provide
the net \hbox{0.5--8~keV} and \hbox{2--8~keV} source counts for the 17 LIRGs in
our sample, and in Figure~3, we show their \hbox{0.5--8~keV}
adaptively-smoothed images.  We note that all galaxies, with the exception of
IC~860, are detected in the full band.  In the hard band, NGC~4418,
CGCG~049$-$057, IC~860, and NGC~23 are undetected.  The LIRGs have net
full-band source counts in the range of \hbox{$\approx$30--$10^5$~counts}; for
the 16 sources that were detected in the full band, \xray\ spectral fitting was
performed.

To model the basic spectral properties of the full-band detected LIRGs in our
sample, we utilized the \xray\ spectral modelling program {\ttfamily xspec}
Version~12.5.1 (Arnaud~1996).  We note that the galaxy-wide and nuclear
spectral properties of the LIRG NGC~1068 have been studied extensively in the
literature by, e.g., Young \etal\ (2001); hereafter, we adopt the complex best-fit
spectral models and resulting properties provided by Young \etal\ (2001).

For each observation of the remaining 15 galaxies with full-band detections, we
employed the {\ttfamily CIAO} tool {\ttfamily specextract} to construct source
and background spectra, response matrix files (RMFs), and properly-weighted
ancillary response files (ARFs) over the source and background regions
described above.  For the 13 full-band detected LIRGs that had $\simgt$300~net
counts (excluding NGC~1068), we utilized $\chi^2$ fitting of the data grouped
in bins with a minimum of 20~counts.  For these galaxies, we assumed a simple
spectral model consisting of Galactic absorption (as given in column~[8] of
Table~1), a Raymond-Smith thermal plasma (from hot gas associated with
star-formation activity) with solar abundances, plus a power-law component
(originating from \xray\ binaries and AGN when present).  Similar composite
fits have been performed successfully for ULIRGs (see, e.g., Franceschini
\etal\ 2003; Ptak \etal\ 2003).  We find that our simple model provides a
reasonable fit to these 13 LIRGs ($\chi^2/\nu =$~0.62--1.35;
median[$\chi^2/\nu$]~=~1.09).  For NGC~3221, we find that the best-fit plasma
component provides only a negligible contribution to the total spectrum; we
therefore fit the \xray\ spectrum using only a power-law model.  For the
remaining two full-band detected LIRGs that had \hbox{$\approx$30--50~net
counts} (NGC~4418 and CGCG~049$-$057) we performed simple power-law fits to the
unbinned spectral data by minimizing the Cash statistic (Cash~1979).  For
NGC~4418, which was observed in two exposures (see Table~2), we performed
joint-spectral fitting of the two observations.

The galaxy-wide best-fit model parameters are provided in Table~3.  Using these
models and the distances provided in column~(7) of Table~1, we computed the
total galaxy-wide \hbox{2--10~keV} power output $L_{\rm HX}^{\rm gal}$.  For
sources that were not detected in the hard band, we computed 3$\sigma$ upper
limits to $L_{\rm HX}^{\rm gal}$ using our best-fit spectral model renormalized
to the 3$\sigma$ upper limit on the \hbox{2--8~keV} count-rate.  In the case of
IC~860, which was not detected in the full or hard bands, we assumed a simple
power-law model with $\Gamma = 2.0$.  Our best estimate values of $L_{\rm
X}^{\rm gal}$ are provided in column~(8) of Table~3.  Using our spectral fits,
we estimate that hot gas emission will provide only a negligible contribution
($<$6.5\% for all LIRGs) to the total \hbox{2--10~keV} emission for our LIRGs.
We therefore expect that LIRGs that do not harbor a luminous AGN (see $\S$~4.2
below) will have $L_{\rm HX}^{\rm gal}$ dominated by \xray\ binary emission.

\subsection{AGN Contribution to Total Galaxy X-ray Emission}

As noted in columns~(13) and (14) of Table~1, \hbox{$\approx$5--7} of the LIRGs
in our sample have been identified to harbor an AGN via optical spectroscopy.
Additional evidence for AGN activity in these LIRGs has been observed in
\spitzer\ infrared spectrograph (IRS) data available from the archive for the
entire sample.  These data were analyzed following the methods outlined in
Goulding \& Alexander~(2009).  The \spitzer\ data reveal emission-line features
(e.g., [Ne~V] $\lambda$14.3$\mu$m or 24.3$\mu$m) indicative of AGN activity for
NGC~1068, NGC~1365, NGC~4194, ESO~420$-$G013, NGC~5135, and NGC~7771, and in
the case of Arp~299 a hot-dust component from an AGN has been identified by
Alonso-Herrero \etal\ (2009).  Furthermore, \xray\ studies of NGC~1068 (e.g.,
Ogle \etal\ 2003), NGC~1365 (e.g., Risaliti \etal\ 2009b), Arp~299 (e.g.,
Della~Ceca \etal\ 2002), NGC~5135 (Levenson \etal\ 2004), and NGC~7771 (Jenkins
\etal\ 2005) have revealed \xray\ spectral signatures of AGN activity
(primarily through the presence of a heavily-absorbed \xray\ emission component
and/or Fe~K emission lines).  In column~(10) of Table~3, we highlight whether a
source has evidence for AGN activity from the optical, infrared, and/or \xray\
bands.

In cases where the AGN is luminous in the \xray\ band, the AGN component will
dominate the integrated (galaxy-wide plus nuclear) \hbox{2--10~keV} power
$L_{\rm HX}^{\rm gal}$ (column~8 of Table~3).  In a forthcoming paper
(Stratford \etal\ in-prep), we will present multiwavelength analyses that
characterize the detailed accretion properties of the AGNs in our LIRG sample;
however, for our purposes, we are primarily interested in how the galaxy-wide
\hbox{2--10~keV} emission from \xray\ binaries correlates with SFR and
$M_\star$ and therefore remove the sources that have \hbox{2--10~keV} emission
dominated by AGN activity.  

To estimate the AGN contribution to $L_{\rm HX}^{\rm gal}$ for each LIRG, we
began by estimating the position of the nucleus of each galaxy.  For the
majority of the LIRGs, the nucleus was isolated first by eye using the DSS red
images and then by choosing manually the {\ttfamily wavdetect} position of the
brightest \hbox{0.5--8~keV} source within the optical nuclear region.  The
adopted LIRG nuclei are generally consistent with the apparent optical nuclei;
their locations are shown in Figures~1 and 3 ({\it gray crosses\/}).  In most
cases, the \xray\ nucleus can be obviously characterized as a bright
point-source at the center of diffuse \xray\ emission or a single point-source
at the center of the optical nuclear region.  However, for Arp~299 and
NGC~5135, there was more than one possible choice for the central nuclear
\xray\ source; for these galaxies, we adopted the positions of the AGN
components noted by Zezas \etal\ (2003) and Levenson \etal\ (2004),
respectively.  Furthermore, for NGC~5653 and IC~860 there were no obvious
central \xray\ point sources to identify as the nucleus; we therefore estimated
by eye the nuclear positions of these galaxies using the DSS red images.

At the nuclear position of each source, we extracted nuclear source counts
$S^{\rm nuc}_{\rm tot}$ and mean exposure times $T^{\rm nuc}_{\rm tot}$ from
the \hbox{2--8~keV} images and exposure maps, respectively, using a circular
aperture with radius equal to the $\approx$90\% encircled-energy fraction
radius (Feigelson \etal\ 2000).\footnote{Feigelson et~al.~(2000) is available
on the WWW at http://www.astro.psu.edu/xray/acis/memos/memoindex.html.}  We
estimated the local nuclear background count-rate for each source by extracting
nuclear background counts $S^{\rm nuc}_{\rm bkg}$ and exposure times $T^{\rm
nuc}_{\rm bkg}$ from an annulus centered on the nuclear position with inner and
outer radii equal to 1.5 and 2.5 times the $\approx$90\% encircled-energy
fraction radius, respectively.  The net nuclear \hbox{2--8~keV} count-rates
\hbox{$\Phi^{\rm nuc}_{\rm 2-8~keV} = S^{\rm nuc}_{\rm tot}/T^{\rm nuc}_{\rm
tot} - S^{\rm nuc}_{\rm bkg}/T_{\rm bkg}^{\rm nuc}$} were then computed and
compared with the total galaxy-wide count-rates $\Phi_{\rm 2-8~keV}^{\rm gal}$
computed in $\S$~4.1.  

In column~(9) of Table~3, we provide $\Phi_{\rm 2-8~keV}^{\rm nuc}$/$\Phi_{\rm
2-8~keV}^{\rm gal}$ for the LIRGs in our sample.  We note that for NGC~1068, we
utilized the nuclear and galaxy-wide \hbox{2--10~keV} luminosities provided by
Young \etal\ (2001).  Sources without listed values of $\Phi_{\rm 2-8~keV}^{\rm
nuc}$/$\Phi_{\rm 2-8~keV}^{\rm gal}$ had both galactic and nuclear components
that were undetected in the \hbox{2--8~keV} band.  For NGC~23, the nucleus was
detected in the \hbox{2--8~keV} band, but due to the increased background over
the larger galaxy-wide area, the total galaxy emission was formally undetected;
for this source, we listed the 3$\sigma$ lower limit on $\Phi_{\rm
2-8~keV}^{\rm nuc}$/$\Phi_{\rm 2-8~keV}^{\rm gal}$.  We consider the observed
\hbox{2--10~keV} emission of the LIRGs to be ``AGN dominant'' if there is both
evidence for AGN activity from the optical, infrared, and/or \xray\ as noted in
column~(10) in Table~3 and $\Phi^{\rm nuc}_{\rm 2-8~keV}/\Phi_{\rm
2-8~keV}^{\rm gal} \ge 0.5$ (see column~[9] of Table~3).  We find that
NGC~1068, NGC~1365, NGC~4194, and NGC~7771 satisfy these conditions.    In
$\S\S$5 and 6 below, we study galaxies with \hbox{2--10~keV} emission dominated
by \xray\ binary populations, and therefore exclude the four AGN-dominant LIRGs
from our further analyses.  We note that a few of the remaining ``normal''
LIRGs in our sample harbor AGNs, which likely provide some contribution to the
nuclear \hbox{2--10~keV} emission.  However, since we generally expect that the
unresolved nucleus (at scales $\simlt$500~pc) will also contain powerful
star-formation regions (e.g., Scoville \etal\ 2000; Alonso-Herrero \etal\
2006a), which will contribute to the total galaxy-wide \xray\ binary emission,
we do not attempt to subtract any AGN component in estimating $L_{\rm HX}^{\rm
gal}$.

%
%%%%%%%%%%%%%%%%%%%%%%%%%%%%%%%%%%%%%%%%%%%%%%%%%%%%%%%%%%%%%%%%%%%%%%%%%%%%%%%%%%
% Figure 4
%%%%%%%%%%%%%%%%%%%%%%%%%%%%%%%%%%%%%%%%%%%%%%%%%%%%%%%%%%%%%%%%%%%%%%%%%%%%%%%%%%
%

\begin{figure}
\figurenum{4}
\centerline{
\includegraphics[width=9cm]{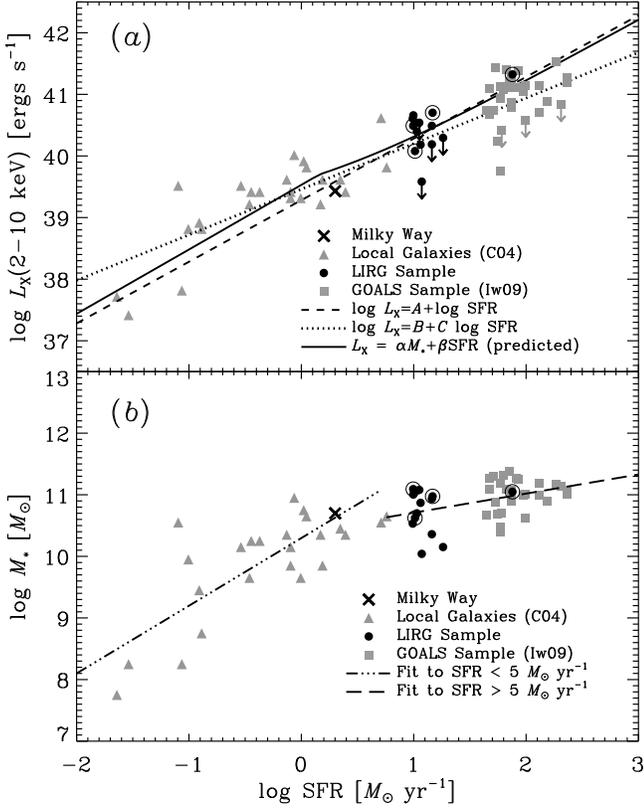}
}
\caption{
({\it a\/}) Galaxy-wide 2--10~keV luminosity $L_{\rm HX}^{\rm gal}$ versus SFR
for the full \chandra\ sample: nearby normal galaxies from C04 ({\it gray
triangles\/}), LIRGs from this study ({\it black circles\/}), and LIRGs/ULIRGs
from the Iw09 GOALS study ({\it gray squares\/}) are plotted.  We have
highlighted ``normal'' LIRGs in our sample that show some evidence for AGN
activity (see column~[10] of Table~3) with open circles.  For reference, we
have plotted the location of the MW as a cross.  We have plotted the mean value
of $\log L_{\rm HX}^{\rm gal}$/SFR as a dashed line and our best-fit relation
assuming a power-law with $\log L_{\rm HX}^{\rm gal} = B + C \log {\rm SFR}$ as
a dotted line.  The expected SFR-variable relation is plotted as a solid curve;
this relation was derived using our best-fit relation for $L_{\rm HX}^{\rm gal}
= \alpha M_\star + \beta {\rm SFR}$ and the relationship between SFR and
$M_\star$ plotted in Fig.~4$b$.  We note that this curve predicts well the
general trend for the \xray\ emission per SFR and tends to decrease with
increasing SFR beyond SFR~$\approx$~1--10~\sfr.
({\it b\/}) $M_\star$ versus SFR for the full \chandra\ sample; symbols have
the same meaning as they did in Fig.~4$a$.  The dot-dashed and long-dashed
curves highlight the best-fit relationship between $M_\star$ and SFR for
galaxies with SFR~$\simlt 5$~\sfr\ and SFR~$\simgt 5$~\sfr, respectively.  We
observe that for nearby galaxies with SFR~$\simlt 5$~\sfr, $M_\star$ and SFR
are roughly linearly proportional; however, at higher SFR values, $M_\star$/SFR
becomes smaller with increasing SFR.
}
\end{figure}
%%%%%%%%%%%%%%%%%%%%%%%%%%%%%%%%%%%%%%%%%%%%%%%%%%%%%%%%%%%%%%%%%%%%%%%%%%%%%%%%%%

%
%%%%%%%%%%%%%%%%%%%%%%%%%%%%%%%%%%%%%%%%%%%%%%%%%%%%%%%%%%%%%%%%%%%%%%%%
\section{Results}
%%%%%%%%%%%%%%%%%%%%%%%%%%%%%%%%%%%%%%%%%%%%%%%%%%%%%%%%%%%%%%%%%%%%%%%%
%

In this section, we explore the correlation between \xray\ luminosity and SFR
using exclusively \chandra\ data from the nearby normal galaxies from C04,
LIRGs from this work, and LIRGs/ULIRGs from the GOALS (Iw09); hereafter, we
refer to these three data sets collectively as the {\it full Chandra sample}.
In the analyses that follow, we have removed galaxies that are observed to
harbor an \xray\ luminous AGN, and for the C04 sample, we removed early-type
galaxies that do not have accurate measures of SFR.  For the C04 sample, AGNs
played a negligible role in the total \xray\ emission from their galaxies and
therefore all 24 late-type galaxies (i.e., spirals and irregulars) are included
here.  For the LIRG sample, the AGNs were identified following the techniques
in $\S$4.2; the remaining 13 ``normal'' LIRGs are members of the full \chandra\
sample.  For the Iw09 sample of 44 LIRGs/ULIRGs, we excluded the 15 obvious
AGNs noted by Iw09.  We note, however, that Iw09 utilized \xray\ hardness
and/or the presence of a strong Fe~K line at $\approx$6.4~keV to identify AGN
activity in their LIRGs/ULIRGs, which is somewhat different from the methods
used for our LIRG sample (see $\S$4.2).  For example, Arp~299, the only galaxy
present in both the Iw09 LIRGs/ULIRGs and our LIRG sample, was classified as an
AGN by Iw09; hereafter, we adopt our ``normal galaxy'' classification and
physical properties for this source.  For the remaining 29 ``normal''
LIRGs/ULIRGs from the Iw09 sample, 23 had estimates of $M_\star$ and SFR from
H10, and for the 6 galaxies without H10 counterparts, we utilized the methods
described in $\S$2.2 to calculate $M_\star$ and SFR.  In total, our full
\chandra\ sample consists of 66 normal galaxies (i.e., 24 from C04, 13 LIRGs
from this study, and 29 LIRGs/ULIRGs from Iw09) and we use these galaxies in
the statistical analyses that follow.

\subsection{A {\itshape Chandra} Perspective on the X-ray/SFR Correlation}

In Figure~4$a$, we show the galaxy-wide \hbox{2--10~keV} luminosity $L_{\rm
HX}^{\rm gal}$ versus SFR for the full \chandra\ sample.  Beginning with a
simple constant $L_{\rm HX}^{\rm gal}$/SFR model, $\log L_{\rm HX}^{\rm
gal}/{\rm SFR} = A$, where $A$ is a constant, we utilized the Kaplan-Meier
estimator (e.g., Feigelson \& Nelson 1986) available through the Astronomy
SURVival Analysis software package ({\ttfamily ASURV} Rev.~1.2; Isobe \&
Feigelson 1990; LaValley \etal 1992) to calculate the mean logarithmic ratio
$A$.  The Kaplan-Meier estimator appropriately handles the censored data (i.e.,
$L_{\rm HX}^{\rm gal}$ upper limits) in our sample.  We find $A = 39.24 \pm
0.06$ (in logarithmic units of ergs~s$^{-1}$~[$M_\odot$~yr$^{-1}$]$^{-1}$;
1$\sigma$ error) with a 1$\sigma$ scatter of 0.48~dex; this relation is plotted
in Fig~4$a$ with a dashed line and is tabulated in Table~4.  Assuming a
power-law scenario where $\log L_{\rm HX}^{\rm gal} = B + C \log {\rm SFR}$, we
utilized the expectation-maximization ({\ttfamily EM}) algorithm available
through {\ttfamily ASURV} to find a best-fit relation of $B = 39.46 \pm 0.06$
and $C = 0.76 \pm 0.04$ (see Fig.~4$a$, {\it dotted curve\/} and Table~4).  We
note that the use of a power-law scenario reduces the scatter by
$\approx$0.1~dex ($\approx$0.39~dex scatter); however, the assumption of such a
highly non-linear relation over the entire SFR range is difficult to explain on
physical grounds.  

We note that at SFR~$\simgt$~1--10~\sfr, the slope of the \xray/SFR relation
appears to become flatter.  Grimm \etal\ (2003) and Gilfanov \etal\ (2004a,b)
argue that such a change in slope is expected in a scenario where the \xray\
emission is dominated by HMXBs and a universal HMXB luminosity function with a
normalization that scales linearly with SFR is applied.  In this scenario, at
\hbox{SFR~$\simlt$~1--10~\sfr}, the relation is expected to be steeper than
linear ($L_{\rm HX}^{\rm gal} \propto$~SFR$^{1.7}$), and at
\hbox{SFR~$\simgt$~1--10~\sfr}, the relation is expected to become linear
($L_{\rm HX}^{\rm gal} \propto$~SFR).  Dividing our data into low and high SFR
regimes around SFR~=~5~\sfr\ and performing power-law fits, we find 
\[L_{\rm HX}^{\rm gal} = \left\{ 
\begin{array}{l r}
  10^{(39.57 \pm 0.11)} {\rm SFR}^{(0.94 \pm 0.15)} & \quad {\rm SFR}~\simlt
0.4~M_\odot~{\rm yr}^{-1} \\
  10^{(39.49 \pm 0.21)} {\rm SFR}^{(0.74 \pm 0.12)} & \quad {\rm SFR}~\simgt
0.4~M_\odot~{\rm yr}^{-1},  \\ \end{array} \right. \]
where the cut-off SFR~$\approx 0.4$~\sfr\ was determined by setting the upper
and lower relations for $L_{\rm HX}^{\rm gal}$ equal to each other and solving
for SFR.  This two-regime relation can be used to obtain a reasonable (with a
1$\sigma$ scatter of $\approx$0.4~dex; see Table~4) estimate of $L_{\rm
HX}^{\rm gal}$ based on SFR alone; however, as we will discuss in $\S$~6.2
below, some LIRGs and ULIRGs may fall $\approx$0.5--1~dex below this relation
due to significant intrinsic attenuation of the \hbox{2--10~keV} emission.  We
note that the measured slopes of these relations are inconsistent with those
predicted by Grimm \etal\ (2003) and Gilfanov \etal\ (2004a,b).  We speculate
that this is due to the presence of LMXBs that provide a non-negligible
contribution to $L_{\rm HX}^{\rm gal}$ in the low-SFR regime.  In the next
section, we explore the use of a physically-motivated scaling of $L_{\rm
HX}^{\rm gal}$ with $M_\star$ and SFR that incorporates emission from both HMXB
and LMXB populations.

\subsection{Correlating $L_{\rm HX}^{\rm gal}$ with $M_\star$ and SFR}

From Figure~4$a$, we notice that the mean $L_{\rm HX}^{\rm gal}$/SFR ratio
({\it dashed line\/}) for the total \chandra\ sample generally lies below the
data for galaxies with SFR~$\simlt$~5~\sfr\ (i.e., from C04), primarily due to
the lower values of $L_{\rm HX}^{\rm gal}$/SFR for the many LIRGs/ULIRGs in the
sample.  This trend has been noted by PR07, who suggest that typical nearby
galaxies (SFR~$\simlt$~5~\sfr) will have fractionally important contributions
from both HMXBs and LMXBs, making them more \xray\ luminous per unit SFR than
``star-formation active'' galaxies like LIRGs/ULIRGs, which are likely to have
negligible contributions from LMXBs and will be dominated by HMXBs.  As
recognized by C04, the total \hbox{2--10~keV} luminosity for nearby galaxies
can reasonably be quantified as the sum of contributions from LMXBs and HMXBs,
which to first order are expected to scale linearly with $M_\star$ and SFR,
respectively.\footnote{However, see Grimm \etal\ (2003), Gilfanov \etal\
(2004a), and Gilfanov~(2004) for different scalings depending on $M_\star$ and
SFR regimes.}  The C04 relation can be expressed using the following equation:
\begin{equation}
L_{\rm HX}^{\rm gal} = L_{\rm HX}^{\rm gal}({\rm LMXB}) + L_{\rm HX}^{\rm
gal}({\rm HMXBs}) = \alpha M_\star + \beta {\rm SFR},
\end{equation}
where $\alpha$ and $\beta$ are scaling constants (see below).  From equation~3,
we expect that galaxies with larger SFR/$M_\star$ will have more significant
contributions from HMXBs, whereas galaxies with smaller SFR/$M_\star$ will have
larger contributions from LMXBs.  We would therefore expect that $L_{\rm
HX}^{\rm gal}$ should scale more closely with SFR/$M_\star$ than SFR alone.  

%
%%%%%%%%%%%%%%%%%%%%%%%%%%%%%%%%%%%%%%%%%%%%%%%%%%%%%%%%%%%%%%%%%%%%%%%%%%%%%%%%%%
% Figure 5
%%%%%%%%%%%%%%%%%%%%%%%%%%%%%%%%%%%%%%%%%%%%%%%%%%%%%%%%%%%%%%%%%%%%%%%%%%%%%%%%%%
%

\begin{figure}
\figurenum{5}
\centerline{
\includegraphics[width=9cm]{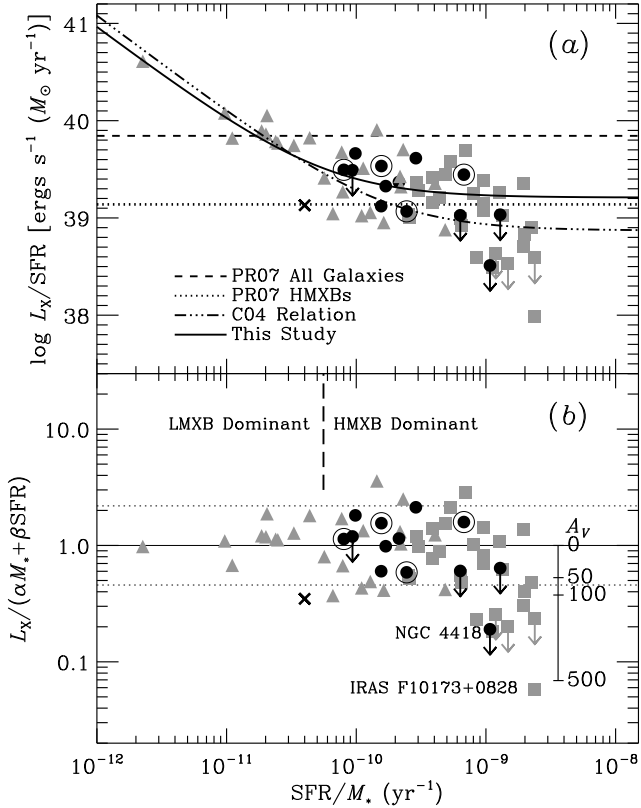}
}
\caption{
({\it a\/}) Logarithm of the galaxy-wide \hbox{2--10~keV} luminosity per SFR
($L_{\rm HX}^{\rm gal}$/SFR) versus SFR/$M_\star$ for the full \chandra\ sample;
symbols are the same as they were in Fig.~4$a$.  The PR07 relations for all
galaxies and HMXBs are shown as dashed and dotted lines, respectively.  The C04
relation and our best-fit relation are shown as dashed-dot and solid curves,
respectively.  The clear dependence of $L_{\rm HX}^{\rm gal}$/SFR on
SFR/$M_\star$ convincingly illustrates that SFR plus $M_\star$ provides a
superior estimate of $L_{\rm HX}^{\rm gal}$ than a simple $L_{\rm HX}^{\rm
gal}$/SFR correlation. 
({\it b\/}) Residuals to the best-fit relation $L_{\rm HX}^{\rm gal}  = \alpha
M_\star + \beta {\rm SFR}$.  Horizontal dotted lines show the 1$\sigma$ scatter
for this relation.  We note that for the most star-formation active galaxies,
there are several sources that are \xray\ underluminous; we have annotated the
locations of NGC~4418 and IRAS~F10173+0828, the most extreme outliers of the
LIRGs and ULIRGs, respectively.  For such sources, it is plausible that
significant extinction ($A_V \simgt 50$~mag), which has been observed in
powerful ULIRGs (see $\S$~6.2), will play a role in obscuring the
\hbox{2--10~keV} emission.  For reference, we have plotted the expected
attenuation of the \hbox{2--10~keV} emission for optical extinction in the
range of \hbox{$A_V =$~0--500~mag} (corresponding to $N_{\rm H}
\approx$~[0--1.2]~$\times 10^{24}$~cm$^{-2}$).
}
\end{figure}
%%%%%%%%%%%%%%%%%%%%%%%%%%%%%%%%%%%%%%%%%%%%%%%%%%%%%%%%%%%%%%%%%%%%%%%%%%%%%%%%%%

In Figure~4$b$, we show $M_\star$ versus SFR for the full \chandra\ sample.  It
is apparent that there is a strong correlation between $M_\star$ and SFR
(Spearman's $\rho$ probability of $>$99.99\%).  Using simple polynomial fitting
techniques, we find for galaxies with \hbox{SFR~$\simlt$~5~\sfr}, $M_\star$ and
SFR are nearly linearly correlated ($M_\star \propto {\rm SFR}^{1.10}$);
however, for galaxies with \hbox{SFR~$\simgt$~5~\sfr}, the relationship becomes
shallower ($M_\star \propto {\rm SFR}^{0.27}$).  From this analysis, we find
that the quantity SFR/$M_\star$ remains roughly constant for
\hbox{SFR~$\approx$~0.01--1~\sfr}, and to first order, this plausibly suggests
that the HMXB and LMXB fractional contributions would remain roughly constant
as well.  Indeed, as noted in $\S$~5.1 above, we found that $L_{\rm HX}^{\rm
gal} \propto {\rm SFR}^{(0.94 \pm 0.15)}$ over this range of SFR.  At
SFR~$\simgt$~5~\sfr, the average SFR/$M_\star$ increases from a constant value,
which would qualitatively allow for HMXBs to make an increasingly large
fractional contribution to the total \xray\ luminosity while the $L_{\rm
HX}^{\rm gal}$/SFR ratio decreases due to the reduced fractional contribution
from LMXBs; in $\S$~5.1 above, we found that $L_{\rm HX}^{\rm gal} \propto {\rm
SFR}^{(0.74 \pm 0.12)}$ in this regime.  

%%%%%%%%%%%%%%%%%%%%%%%%%%%%%%%%%%%%%%%%%%%%%%%%%%%%%%%%%%%%%%%%%%%%%%%%%%%%%%%%%%
% Table 4
%%%%%%%%%%%%%%%%%%%%%%%%%%%%%%%%%%%%%%%%%%%%%%%%%%%%%%%%%%%%%%%%%%%%%%%%%%%%%%%%%%
\begin{table*}
\begin{center}
\caption{Best-Fit \xray/SFR Relations}
\begin{tabular}{lcc}
\hline\hline
\multicolumn{1}{c}{\sc Model Description} & \multicolumn{1}{c}{\sc Best-Fit Parameters} & \multicolumn{1}{c}{1$\sigma$ {\sc Scatter} (dex)} \\
\hline\hline
$\log L_{\rm HX}^{\rm gal} = A + \log {\rm SFR}$ \ldots\ldots\ldots\ldots\ldots\ldots\ldots\ldots\ldots\ldots\ldots\ldots\ldots\ldots\ldots & $A = 39.24 \pm 0.06$ & 0.48 \\ 
$\log L_{\rm HX}^{\rm gal} = B + C \log {\rm SFR}$ \dotfill & $B = 39.46 \pm 0.06$ and $C = 0.76 \pm 0.04$ & 0.39 \\ 
$\log L_{\rm HX}^{\rm gal} = B + C \log {\rm SFR}$ (SFR~$\simlt 0.4$~\sfr) \dotfill & $B = 39.57 \pm 0.11$ and $C = 0.94 \pm 0.15$ & 0.41 \\ 
$\log L_{\rm HX}^{\rm gal} = B + C \log {\rm SFR}$ (SFR~$\simgt 0.4$~\sfr) \dotfill & $B = 39.49 \pm 0.21$ and $C = 0.74 \pm 0.12$ & 0.36 \\ 
$L_{\rm HX}^{\rm gal} = \alpha M_\star + \beta {\rm SFR}$  \dotfill & $\alpha = (9.05 \pm 0.37) \times 10^{28}$~ergs~s$^{-1}$~$M_\odot^{-1}$ & \\
 & $\beta = (1.62 \pm 0.22) \times 10^{39}$~ergs~s$^{-1}$~($M_\odot$~yr$^{-1}$)$^{-1}$  & 0.34 \\ 
\hline
\end{tabular}
\end{center}
NOTE.---Summary of best-fit relations between $L_{\rm HX}^{\rm gal}$, SFR, and $M_\star$ for the models described in $\S$~5.  Fits have been performed following the methods described in $\S\S$5.1 and 5.2; all quoted errors are 1$\sigma$.  We note that the inclusion of both SFR and $M_\star$ in the final model provides the least statistical scatter of all the models.
\end{table*}

%%%%%%%%%%%%%%%%%%%%%%%%%%%%%%%%%%%%%%%%%%%%%%%%%%%%%%%%%%%%%%%%%%%%%%%%%%%%%%%%%%

The best estimates to date of the quantities $\alpha$ and $\beta$ come from the
C04 analysis, which does not include \chandra\ constraints on galaxies with
SFR~$\simgt$~5~\sfr\ (i.e., LIRGs and ULIRGs), a regime where the SFR/$M_\star$
is expected to be large and HMXBs are likely to dominate.  Here, we utilize the
full \chandra\ sample to improve the calibration of $\alpha$ and $\beta$ for
data covering the broad range of SFR~$\approx$~0.01--400~\sfr.

Considering the above discussion, it is useful to examine the relationship of
$L_{\rm HX}^{\rm gal}$/SFR as a function of SFR/$M_\star$.  Following
equation~3, the quantity $L_{\rm HX}^{\rm gal}$/SFR can be written as follows:
\begin{equation}
L_{\rm HX}^{\rm gal}/{\rm SFR} = (\alpha M_\star + \beta {\rm SFR})/{\rm SFR} =
\alpha ({\rm SFR}/M_\star)^{-1} + \beta.
\end{equation}
From equation~4, we see that $L_{\rm HX}^{\rm gal}$/SFR~$\propto ({\rm
SFR}/M_\star)^{-1}$, and in Figure~5$a$, we plot $L_{\rm HX}^{\rm gal}$/SFR
versus SFR/$M_\star$ for the full \chandra\ sample.  We see that the IMF and
\xray\ bandpass adjusted C04 relation ({\it dot-dashed} curve in Fig~5$a$),
which was determined solely from the C04 data ({\it triangles\/}), predicts
well the basic observed trend; however, it mildly underpredicts $L_{\rm
HX}^{\rm gal}$/SFR for LIRGs and ULIRGs.  Using the full \chandra\ sample, we
fit the data using the {\ttfamily EM} algorithm and the model provided in
equation~4.  We find best-fit values of $\alpha = (9.05 \pm 0.37) \times
10^{28}$~ergs~s$^{-1}$~$M_\odot^{-1}$ and $\beta = (1.62 \pm 0.22) \times
10^{39}$~ergs~s$^{-1}$~($M_\odot$~yr$^{-1}$)$^{-1}$ with a scatter of 0.34~dex,
which is smaller than that produced by fitting for SFR alone (see $\S$~5.1;
Table~4).  

Our best-fit relation is plotted as a solid curve in Figure~5$a$.  In
Figure~4$a$, we combine this relation with the \hbox{$M_\star$--SFR} relations
provided in $\S$~5.1 and Figure~4$b$ to show the expected \hbox{$L_{\rm
HX}^{\rm gal}$--SFR} relation (see {\it solid curve} in Fig.~4$a$); in general,
the combined relations appear to fit the data better than a simple mean $\log
L_{\rm HX}^{\rm gal}$/SFR ratio and a less physically meaningful power-law
relation (i.e., $\log L_{\rm HX}^{\rm gal} = B + C {\rm SFR}$).  

After accounting for differences in IMF and \xray\ bandpass, we find that
$\alpha$/$\alpha$(C04)~=~0.76 and $\beta$/$\beta$(C04)~=~2.18, suggesting that
for a given SFR/$M_\star$, HMXBs provide $\approx$2.9 times larger $L_{\rm
X}^{\rm gal}$ contributions over LMXBs than previously reported.  We further
note that our best-fit relation has reduced the error bars for $\alpha$ and
$\beta$ by factors of $\approx$3.8 and $\approx$2.1, respectively.  We find
that for SFR/$M_\star$~$\simgt$~$5.6 \times 10^{-11}$~yr$^{-1}$ (corresponding
roughly to SFR $\simgt 2$~\sfr\ or $L_{\rm IR} \simgt 2 \times 10^{10}
L_{\odot}$ assuming the $M_\star$--SFR correlation in Fig.~4$b$ and $\gamma =1$
in equation~2), HMXBs will dominate the galaxy-wide \hbox{2--10~keV}.  This
transition SFR/$M_\star$ is a factor of $\approx$2.9 times lower than previous
estimates.

We note that the MW (plotted in Fig.~5 as a {\it cross\/}) appears to be
located below our best-fit correlation line by a factor of $\approx$0.37~dex,
which is just outside the 1$\sigma$ scatter in the relation.  Our best-fit
relation predicts that a galaxy with $M_\star$ and SFR equal to those of the
MW, will have $L_{\rm HX}^{\rm gal}$(LMXB)/$L_{\rm HX}^{\rm
gal}$(HMXB)~$\approx 1.4$, which is much less than the $L_{\rm HX}^{\rm
gal}$(LMXB)/$L_{\rm HX}^{\rm gal}$(HMXB)~$\sim 10$ estimated by Grimm \etal\
(2002) using the measured luminosity functions of LMXBs and HMXBs.  We
speculate that these differences are due to either (1) the large uncertainties
in measuring $M_\star$ and SFR for the MW, as a result of our location within
it, and/or (2) the MW being a true outlier to the general correlation.  It has
been noted by Hammer \etal\ (2007), that the MW does not appear to follow other
correlations along physical parameter planes (e.g., planes drawn by disk
rotation velocity, $K$-band luminosity, disk angular momentum, and mean [Fe/H]
abundances of stars in the outskirts of the disk) drawn using a sample of
galaxies with similar $M_\star$.  Hammer \etal\ conclude that the MW is likely
to have a different star-formation history and disk construction than other
similar galaxies; such conditions could plausibly affect the scaling of LMXB
and HMXB power output with $M_\star$ and SFR, respectively.

%
%%%%%%%%%%%%%%%%%%%%%%%%%%%%%%%%%%%%%%%%%%%%%%%%%%%%%%%%%%%%%%%%%%%%%%%%
\section{Discussion}
%%%%%%%%%%%%%%%%%%%%%%%%%%%%%%%%%%%%%%%%%%%%%%%%%%%%%%%%%%%%%%%%%%%%%%%%
%

In Figure~5$b$, we show the residuals to our best-fit relation (i.e., $L_{\rm
X}/[\alpha M_\star + \beta {\rm SFR}]$).  As noted in $\S$~5.2, the scatter
over the entire range of SFR/$M_\star$ is $\approx$0.34~dex.  There are several
factors that are likely to contribute to the scatter in this relation including:
(1) variations in the LMXB populations due to a diversity of stellar ages and
star-formation histories (e.g., Fragos \etal\ 2008); (2) unrelated low-level AGN
activity that affect estimates of the \xray\ and/or infrared luminosities, two
quantities that are important to this analysis (e.g., Imanishi \etal\ 2007,
2010; Iw09); (3) strong statistical fluctuations due to small numbers of
luminous \xray\ binaries that dominate the total \xray\ luminosity,
particularly for galaxies with small SFR and/or $M_\star$ (e.g., Grimm \etal\
2003; Gilfanov \etal\ 2004a; see $\S$~6.1 below); and (4) significant
extinction that can affect estimates of stellar mass, and in the most extreme
cases, may obscure even \hbox{2--10~keV} emission (see $\S$~6.2 below).  In
$\S\S$~6.1 and 6.2 below, we explore in more detail points (3) and (4),
respectively.

\subsection{Statistical Considerations at Low SFR and $M_\star$}

As discussed in $\S$~5.1, the qualitative change in slope of the \xray/SFR
correlation has previously been noted by Grimm \etal\ (2003) and Gilfanov
\etal\ (2004a,b) who predict such behavior for an HMXB-dominant $L_{\rm
HX}^{\rm gal}$.  These authors suggest that since the HMXB luminosity function
is a relatively shallow power-law with $dN/dL_{\rm XP} \propto {\rm SFR} \;
L_{\rm XP}^{-\gamma}$ (with $\gamma \approx$~1.6) and a maximum cut-off
luminosity of $L_{\rm XP}^{\rm cut-off} \approx 2 \times 10^{40}$~\xlum\ (see,
e.g., Grimm \etal\ 2003; Swartz \etal\ 2004), then in the low-SFR regime, the
most luminous HMXB will have a most probable luminosity $L_{\rm XP}^{\rm max}
\propto {\rm SFR}^{1/(\gamma-1)}$, which is less than $L_{\rm XP}^{\rm
cut-off}$.  In this regime, the integrated \xray\ luminosity $L_{\rm HX}^{\rm
gal}$ is therefore expected to scale with ${\rm SFR}^{1/(\gamma-1)}$ (or
SFR$^{1.7}$ in the case of $\gamma = 1.6$).  Such an effect is predicted to
produce a non-linear SFR dependence for the HMXB contribution to $L_{\rm
HX}^{\rm gal}$ for \hbox{SFR~$\simlt$~1--10~\sfr}.  Similar logic can be
applied to the LMXB contribution to $L_{\rm HX}^{\rm gal}$ using the more
complex luminosity function for LMXBs, which is expected to have a linear
normalization that scales with $M_\star$.  However, the cut-off luminosity for
the LMXB population is likely to be much lower ($\sim$10$^{39}$~\xlum) than
that of HMXBs and therefore $L_{\rm HX}^{\rm gal}$(LMXB) is expected to be
linear for galaxies with $M_\star \simgt$~5--20~$\times 10^9$~$M_\odot$ and
very close to linear for smaller values of $M_\star$ (e.g., Gilfanov \etal\
2004a; Gilfanov~2004).  

%
%%%%%%%%%%%%%%%%%%%%%%%%%%%%%%%%%%%%%%%%%%%%%%%%%%%%%%%%%%%%%%%%%%%%%%%%%%%%%%%%%%
% Figure 6
%%%%%%%%%%%%%%%%%%%%%%%%%%%%%%%%%%%%%%%%%%%%%%%%%%%%%%%%%%%%%%%%%%%%%%%%%%%%%%%%%%
%

\begin{figure*}
\figurenum{6}
\centerline{
\includegraphics[width=19cm]{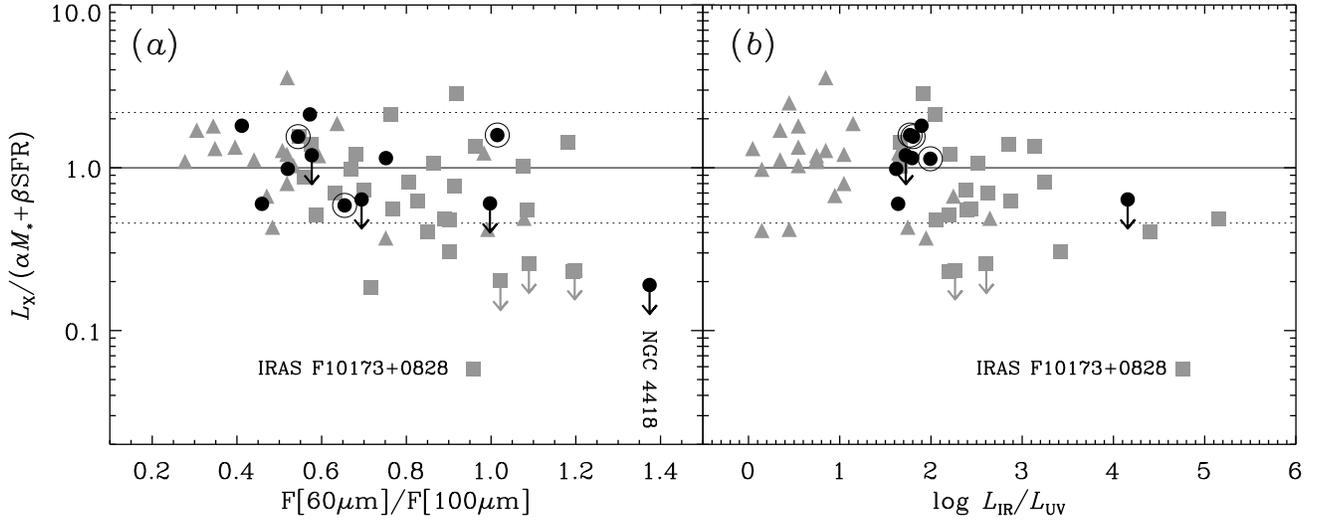}
}
\caption{
Residuals to the best-fit relation $L_{\rm HX}^{\rm gal}  = \alpha M_\star + \beta {\rm
SFR}$ versus the IRAS 60--to--100$\mu$m flux density ratio $F_{60}/F_{100}$
({\it a\/}) and the logarithm of the infrared-to-UV luminosity ratio $\log
L_{\rm IR}/L_{\rm UV}$ ({\it b\/}).  Symbols and lines have the same meaning as
they did in Figure~5$b$.  A Spearman's $\rho$ test reveals that the residuals
are negatively correlated with $F_{60}/F_{100}$ and $\log L_{\rm IR}/L_{\rm
UV}$ at the $\approx$99.8\% and 99.9\% confidence levels, respectively.  The
quantity $F_{60}/F_{100}$ is a proxy for temperature and infrared luminosity
density (star-formation density) and $\log L_{\rm IR}/L_{\rm UV}$ is a direct
measure of UV obscuration.  These relations suggest that the most \xray\
deficient LIRGs and ULIRGs, which have relatively warm dust temperatures and
large UV obscuration due to dust, contain dense buried star-forming regions
with attenuated \hbox{2--10~keV} emission.
}
\end{figure*}
%%%%%%%%%%%%%%%%%%%%%%%%%%%%%%%%%%%%%%%%%%%%%%%%%%%%%%%%%%%%%%%%%%%%%%%%%%%%%%%%%%

In $\S$~5.1, we noted that in the low-SFR regime (SFR~$\approx$
\hbox{0.01--1~\sfr}), $L_{\rm HX}^{\rm gal}$/SFR and SFR/$M_\star$ were roughly
constant; over this range these quantities have median values of $\approx$$5
\times 10^{39}$~\xlum~($M_\odot$~yr$^{-1}$)$^{-1}$ and $8 \times
10^{-11}$~yr$^{-1}$, respectively.  From our best-fit relation for equation~3,
we estimate $L_{\rm HX}^{\rm gal}$(LMXB)/$L_{\rm HX}^{\rm gal}$(HMXB)~$\approx
0.74$ for these galaxies, suggesting that both LMXBs and HMXBs are estimated to
provide significant contributions to $L_{\rm HX}^{\rm gal}$ in this regime.
However, if indeed $L_{\rm HX}^{\rm gal}({\rm HMXB}) \propto {\rm SFR}^{1.7}$
and $L_{\rm HX}^{\rm gal}({\rm LMXB}) \propto M_\star \propto {\rm SFR}$ are
accurate descriptions of the respective HMXB and LMXB emission in this regime,
then we would have expected that $L_{\rm HX}^{\rm gal} \propto {\rm
SFR}^{1.0-1.7}$.  In $\S$~5.1, we found that $L_{\rm HX}^{\rm gal} \propto {\rm
SFR}^{(0.94 \pm 0.15)}$.  This suggests that either (1) $L_{\rm HX}^{\rm
gal}({\rm HMXB})$ scales with ${\rm SFR}^{1.7}$ and provides a small
contribution to $L_{\rm HX}^{\rm gal}$ or (2) $L_{\rm HX}^{\rm gal}({\rm HMXB})
\propto {\rm SFR}$.  Future \xray\ and multiwavelength studies of large numbers
of normal galaxies in this regime may provide the necessary statistical basis
to allow discrimination between these two scenarios; future \xray\ missions
such as \ixo\ and \wfxt\ will enable efficient detection of many galaxies in
the low-SFR regime.\footnote{More information regarding \ixo\ and \wfxt\ can be
found at {\ttfamily http://ixo.gsfc.nasa.gov/} and {\ttfamily
http://wfxt.pha.jhu.edu/}, respectively.} Otherwise, a combination of \xray\
and optical/near-IR observatories with highly-precise astrometry would be
needed to allow for the detection of \xray\ binaries and the reliable
identification and classification of their stellar counterparts in external
galaxies.  This is currently broadly possible with the combination of \chandra\
and \hst\ data for a small number of nearby galaxies.  Future missions such as
\genx\ and \jwst\ will undoubtedly expand this science to more representative
galaxies that cover the entire low-SFR regime.\footnote{More information
regarding \genx\ and \jwst\ can be found at {\ttfamily
http://www.cfa.harvard.edu/hea/genx/} and {\ttfamily
http://www.jwst.nasa.gov/}, respectively.}

\subsection{A Deficit of X-ray Emission in the Most Actively Star-Forming Galaxies}

From Figure~5$b$, it is evident that for several of the most star-formation
active LIRGs/ULIRGs (i.e., SFR/$M_\star$~$\simgt$ 10$^{-9}$ yr$^{-1}$) in the
total \chandra\ sample there is a departure from the best-fit relation.  For
example, \hbox{IRAS~07251$-$0248}, \hbox{IRAS~F15250+3608},
\hbox{IRAS~F10173+0828}, \hbox{IRAS~21101+5810}, and NGC~4418 all have
residuals to our best fit that fall below the 2$\sigma$ scatter, when only
\hbox{$\approx$1--2} galaxies are expected statistically over the entire
SFR/$M_\star$ range.  If we exclude galaxies with
\hbox{SFR/$M_\star$~$\simgt$~10$^{-9}$~yr$^{-1}$}, this significantly reduces
the scatter to $\approx$0.26~dex (i.e., by $\approx$20\%).  Studies of
LIRGs/ULIRGs have found that in these objects the powerful star-formation
activity occurs in compact regions that are surrounded by large columns of gas
and dust (e.g., Soifer \etal\ 2000; Alonso-Herrero \etal\ 2006b; Siebenmorgen
\etal\ 2008).  The obscuration to these star-forming regions can be very
significant ($A_V \simgt 50$~mag; e.g., Genzel \etal\ 1998; Siebenmorgen \&
Kr{\"u}gel~2007) and can attenuate emission over the \hbox{2--10~keV} bandpass.  

To estimate the effect that obscuration can have on the observed
\hbox{2--10~keV} power for ULIRG levels of extinction, we first converted $A_V$
to $N_{\rm H}$ following the relation \hbox{$N_{\rm H} \approx 2.3 \times
10^{21} A_V$~cm$^{-2}$} from G{\"u}ver \& {\"O}zel~(2009), which was calibrated
using measurements of various columns through the Milky~Way.  We note that this
calibration has been calibrated only over the range of $A_V
\approx$~0.2--30~mag and is subject to large uncertainties when applied to
other galaxies, which may have different chemical compositions in their
interstellar mediums.  A range of \hbox{$A_V =$~0--500} therefore implies
column densities in the range $N_{\rm H} \approx$ \hbox{0--$1.2 \times
10^{24}$~cm$^{-2}$}.  Using PIMMS and an assumed power-law \xray\ SED with
$\Gamma = 1.8$, we find that the emergent \hbox{2--10~keV} luminosity will be
attenuated by factors of $\approx$2, 3, and 15 for $A_V \approx$~50, 100, and
500, respectively (see vertical bar in Fig.~5$b$); such visual extinctions are
likely to be present in at least some LIRGs/ULIRGs (e.g., Genzel \etal\ 1998
estimate that some ULIRGs may even reach $A_V \approx 1000$).  We note that
there are only a few LIRGs/ULIRGs in our sample with the high-quality optical
to far-IR data required to estimate $A_V$ accurately.  For example, the
well-studied ULIRG Arp~220 has a \hbox{2--10~keV} luminosity that is a factor of
$\approx$3 times lower than that expected from our best-fit relation for
equation~4.  Detailed modelling of the infrared-to-optical spectrum suggests a
global value of \hbox{$A_V \approx$~50--150~mag} (e.g., Siebenmorgen \&
Kr{\"u}gel~2007), implying \hbox{$N_{\rm H} \approx$~1.2--3.5~$\times
10^{23}$~cm$^{-2}$} and therefore an expected factor of \hbox{$\approx$2--4}
attenuation of the intrinsic \hbox{2--10~keV} luminosity, in good agreement
with the deficit that we observe.

To assess whether there is broader evidence that the \xray\ underluminous
LIRGs/ULIRGs in the full \chandra\ sample may be subject to significant
\hbox{2--10~keV} attenuation, we utilized two informative quantities: (1) the
\iras\ 60--to--100$\mu$m flux ratio ($F_{60}/F_{100}$), which has been shown to
correlate well with mean dust temperature and infrared luminosity surface
density (e.g., Chanial \etal\ 2007); and (2) the IR-to-UV luminosity ratio
($L_{\rm IR}/L_{\rm UV}$), which provides a direct measure of UV obscuration
from star-forming regions.  If the \xray\ emission from star-forming regions is
buried under significant absorbing columns, we would expect the star-formation
to be highly concentrated and not dispersed through the galaxy (i.e., large
$F_{60}/F_{100}$) and heavily obscured in the UV (i.e., large $L_{\rm
IR}/L_{\rm UV}$).  In Figure~6$a$ and 6$b$, we plot the residuals to our
best-fit relation ($L_{\rm HX}^{\rm gal}/[\alpha M_\star + \beta{\rm SFR}$])
versus $F_{60}/F_{100}$ and $L_{\rm IR}/L_{\rm UV}$, respectively.  It is
evident that the LIRGs/ULIRGs with the largest deficit of \xray\ emission with
respect to our best-fit relation are those with large values of
$F_{60}/F_{100}$ and $L_{\rm IR}/L_{\rm UV}$; a Spearman's $\rho$ test reveals
that the residuals are correlated with each quantity at the 99.8\% and 99.9\%
significance levels, respectively.  We therefore conclude that obscuration is
at least partially responsible for the deficit of observed \hbox{2--10~keV}
emission at the largest SFR/$M_\star$.

\section{Conclusions}

In this paper, we present new results from \chandra\ observations of a complete
sample of 17 LIRGs with $D < 60$~Mpc and Galactic column densities $N_{\rm H}
\simlt 5 \times 10^{20}$~cm$^{-2}$.  Our \chandra\ observations reveal a
variety of \xray\ morphologies for the LIRGs, with the majority of their
corresponding \xray\ spectra being well-described by a hot-gas component (with
$kT \simlt 0.8$~keV) plus a power-law ($\Gamma \approx$~1--3), which is
expected to originate from \xray\ binaries, and when present, an AGN.  We find
that four out of the 17 LIRGs both show evidence for AGN activity in the
optical, infrared, and/or \xray\ band and contain a nuclear component that
dominates the integrated \hbox{2--10~keV} luminosity.  We study the
galaxy-wide \hbox{2--10~keV} luminosity $L_{\rm HX}^{\rm gal}$ and its
correlation with SFR and $M_\star$ using the 13 non-AGN-dominant LIRGs combined
with \chandra\ studies of less star-formation active normal galaxies (from C04)
and more powerful LIRGs and ULIRGs (from Iw09), collectively referred to as the
total \chandra\ sample.  Our key results can be summarized as follows:\\

\noindent (1)---For the galaxies that make up our total \chandra\ sample, we
find that SFR and $M_\star$ are strongly correlated, such that $M_\star \propto
{\rm SFR}^{1.10}$ for SFR~$\approx$ \hbox{0.01--5}~\sfr\ and 
$M_\star \propto {\rm SFR}^{0.27}$ for SFR~$\simgt$~5~\sfr\ (see Fig.~4$b$).

\noindent (2)---We find that $L_{\rm HX}^{\rm gal}$ is strongly correlated with
SFR (the \xray/SFR correlation), as has been found in previous studies (see
Fig.~4$b$).  Similar to point~(1), the \xray/SFR correlation slope becomes
shallower for SFR~$\simgt$~1--10~\sfr.  We find that $L_{\rm HX}^{\rm gal} =
(39.57 \pm 0.11) {\rm SFR}^{(0.94 \pm 0.15)}$ for SFR~$\simlt$~0.4~\sfr\ and
$L_{\rm HX}^{\rm gal} = (39.49 \pm 0.21) {\rm SFR}^{(0.74 \pm 0.12)}$ for
SFR~$\simgt$~0.4~\sfr\ provides a reasonable (with $\approx$0.4~dex scatter; see Table~4)
estimate of the $L_{\rm HX}^{\rm gal}$.  The change in slope is likely to be
related to the changing slope of the correlation between SFR and $M_\star$,
since we expect the total galaxy-wide \hbox{2--10~keV} luminosity to be the sum
of contributions from HMXBs and LMXBs, which to first order are expected to
scale with SFR and $M_\star$, respectively.

\noindent (3)---Under the assumption that $L_{\rm HX}^{\rm gal}$ can be written
as $L_{\rm HX}^{\rm gal} = L_{\rm HX}^{\rm gal}({\rm LMXB}) + L_{\rm HX}^{\rm
gal}({\rm HMXB}) = \alpha M_\star + \beta {\rm SFR}$, we find best-fitting
parameters of $\alpha = (9.05 \pm 0.37) \times
10^{28}$~ergs~s$^{-1}$~$M_\odot^{-1}$ and $\beta = (1.62 \pm 0.22) \times
10^{39}$~ergs~s$^{-1}$~($M_\odot$~yr$^{-1}$)$^{-1}$.  These values suggest that
HMXBs dominate the \xray\ emission for SFR/$M_\star$~$\simgt 5.9\times
10^{-11}$~yr$^{-1}$, a factor of $\approx$2.9 times lower than previous
estimates.

\noindent (4)---For the most actively star-forming LIRGs and ULIRGs (i.e.,
those with SFR/$M_\star$~$\simgt 10^{-9}$~yr$^{-1})$, we find an excess of
objects with $L_{\rm HX}^{\rm gal}$ lower than expected.  These sources have
infrared colors and UV to total-IR luminosities indicative of compact
star-forming regions with extreme extinction.  We argue that such star-forming
regions are likely to have attenuated \hbox{2--10~keV} emission.

\acknowledgements

We thank the referee for their helpful comments that have improved the
manuscript.  We thank Kazushi Iwasawa for generously sharing data and Lee Armus
and Joseph Mazzarella for helpful discussions.  We gratefully acknowledge
financial support from the Einstein Fellowship Program (B.D.L.), the Royal
Society (D.M.A.), the Leverhulme Trust (D.M.A.), \chandra\ \xray\ Center grant
G09-0134A/B (W.N.B.,F.E.B.), and NASA ADP grant NNX10AC99G (W.N.B.).  This
research has made use of the NASA/IPAC Extragalactic Database (NED), which is
operated by the Jet Propulsion Laboratory, California Institute of Technology,
under contract with NASA.  

%
%%%%%%%%%%%%%%%%%%%%%%%%%%%%%%%%%%%%%%%%%%%%%%%%%%%%%%%%%%%%%%%%%%%%%%%%

%%%%%%%%%%%%%%%%%%%%%%%%%%%%%%%%%%%%%%%%%%%%%%%%%%%%%%%%%%%%%%%%%%%%%%%%%%%%%%%%%%
%


\begin{thebibliography}{}
%%%%%%%%%%%%%%%%%%%%%%%%%%%%%%%%%%%%%%%%%%%%%%%%%%%%%%%%%%%%%%%%%%%%%%%%
%


\bibitem[Alonso-Herrero et al.(2006a)]{2006ApJ...650..835A} Alonso-Herrero, 
A., Rieke, G.~H., Rieke, M.~J., Colina, L., P{\'e}rez-Gonz{\'a}lez, P.~G., 
\& Ryder, S.~D.\ 2006a, \apj, 650, 835 

\bibitem[Alonso-Herrero et al.(2006b)]{2006ApJ...652L..83A} Alonso-Herrero, 
A., Colina, L., Packham, C., D{\'{\i}}az-Santos, T., Rieke, G.~H., 
Radomski, J.~T., \& Telesco, C.~M.\ 2006b, \apjl, 652, L83

\bibitem[Alonso-Herrero et al.(2009)]{2009ApJ...697..660A} Alonso-Herrero, 
A., et al.\ 2009, \apj, 697, 660 

\bibitem[Armus et al.(1989)]{1989ApJ...347..727A} Armus, L., Heckman,
T.~M., \& Miley, G.~K.\ 1989, \apj, 347, 727


\bibitem[Armus et al.(2009)]{2009PASP..121..559A} Armus, L., et al.\ 2009, 
\pasp, 121, 559

\bibitem[Arnaud(1996)]{1996ASP....101} Arnaud, K.A., 1996, Astronomical Data
Analysis Software and Systems V, eds. Jacoby G. and Barnes J., p17, ASP Conf.
Series volume 101

\bibitem[Baan et al.(1998)]{1998ApJ...509..633B} Baan, W.~A., Salzer,
J.~J., \& Lewinter, R.~D.\ 1998, \apj, 509, 633

\bibitem[Bauer et al.(2002)]{2002AJ....124.2351B} Bauer, F.~E., Alexander,
D.~M., Brandt, W.~N., Hornschemeier, A.~E., Vignali, C., Garmire, G.~P., \&
Schneider, D.~P.\ 2002, \aj, 124, 2351

\bibitem[Bauer et al.(2004)]{2004AJ....128.2048B} Bauer, F.~E., Alexander, 
D.~M., Brandt, W.~N., Schneider, D.~P., Treister, E., Hornschemeier, A.~E., 
\& Garmire, G.~P.\ 2004, \aj, 128, 2048 

\bibitem[Bell et al.(2003)]{2003ApJS..149..289B} Bell, E.~F., McIntosh, 
D.~H., Katz, N., \& Weinberg, M.~D.\ 2003, \apjs, 149, 289 

\bibitem[Bell et al.(2005)]{2005ApJ...625...23B} Bell, E.~F., et al.\ 2005, 
\apj, 625, 23 

\bibitem[Cash(1979)]{1979ApJ...228..939C} Cash, W.\ 1979, \apj, 228, 939 

\bibitem[Chen et al.(2010)]{2010ApJ...712.1385C} Chen, Y., Lowenthal, 
J.~D., \& Yun, M.~S.\ 2010, \apj, 712, 1385 

\bibitem[Chanial et al.(2007)]{2007A&A...462...81C} Chanial, P., Flores, H.,
Guiderdoni, B., Elbaz, D., Hammer, F., \& Vigroux, L.\ 2007, \aap, 462, 81

\bibitem[Colbert et al.(2004)]{2004ApJ...602..231C} Colbert, E.~J.~M.,
Heckman, T.~M., Ptak, A.~F., Strickland, D.~K., \& Weaver, K.~A.\ 2004,
\apj, 602, 231 (C04)

\bibitem[Coziol et al.(1998)]{1998ApJS..119..239C} Coziol, R., Torres,
C.~A.~O., Quast, G.~R., Contini, T., \& Davoust, E.\ 1998, \apjs, 119, 239

\bibitem[Della Ceca et al.(2002)]{2002ApJ...581L...9D} Della Ceca, R., et 
al.\ 2002, \apjl, 581, L9 

\bibitem[de Vaucouleurs et al.(1991)]{1991trcb.book.....D} de Vaucouleurs,
G., de Vaucouleurs, A., Corwin, H.~G., Buta, R.~J., Paturel, G., \& Fouque,
P.\ 1991, Volume 1-3, XII, 2069 pp.~7 figs..~ Springer-Verlag Berlin
Heidelberg New York

\bibitem[Dickey \& Lockman(1990)]{1990ARA&A..28..215D} Dickey, J.~M., \&
Lockman, F.~J.\ 1990, \araa, 28, 215 

\bibitem[Fabbiano(1989)]{1989ARA&A..27...87F} Fabbiano, G.\ 1989, \araa,
27, 87

\bibitem[Fabbiano(2006)]{2006ARA&A..44..323F} Fabbiano, G.\ 2006, \araa, 
44, 323 

\bibitem[Feigelson \& Nelson(1985)]{1985ApJ...293..192F} Feigelson, E.~D.,
\& Nelson, P.~I.\ 1985, \apj, 293, 192

\bibitem[Fragos et al.(2008)]{2008ApJ...683..346F} Fragos, T., et al.\ 
2008, \apj, 683, 346 

\bibitem[Franceschini et al.(2003)]{2003MNRAS.343.1181F} Franceschini, A., 
et al.\ 2003, \mnras, 343, 1181 

\bibitem[Garc{\'{\i}}a-Mar{\'{\i}}n et al.(2006)]{2006ApJ...650..850G} 
Garc{\'{\i}}a-Mar{\'{\i}}n, M., Colina, L., Arribas, S., Alonso-Herrero, 
A., \& Mediavilla, E.\ 2006, \apj, 650, 850

\bibitem[Genzel et al.(1998)]{1998ApJ...498..579G} Genzel, R., et al.\ 
1998, \apj, 498, 579 

\bibitem[Gilfanov(2004)]{2004MNRAS.349..146G} Gilfanov, M.\ 2004, \mnras, 
349, 146 

\bibitem[Gilfanov et al.(2004)]{2004MNRAS.351.1365G} Gilfanov, M., Grimm, 
H.-J., \& Sunyaev, R.\ 2004a, \mnras, 351, 1365 

\bibitem[Gilfanov et al.(2004)]{2004NuPhS.132..369G} Gilfanov, M., Grimm,
H.-J., \& Sunyaev, R.\ 2004b, Nuclear Physics B Proceedings Supplements,
132, 369

\bibitem[Goulding \& Alexander(2009)]{2009MNRAS.398.1165G} Goulding, A.~D., \&
Alexander, D.~M.\ 2009, \mnras, 398, 1165

\bibitem[Grimm et al.(2002)]{2002A&A...391..923G} Grimm, H.-J., Gilfanov, M.,
\& Sunyaev, R.\ 2002, \aap, 391, 923 

\bibitem[Grimm et al.(2003)]{2003MNRAS.339..793G} Grimm, H.-J., Gilfanov,
M., \& Sunyaev, R.\ 2003, \mnras, 339, 793

\bibitem[G{\"u}ver {\"O}zel(2009)]{2009MNRAS.400.2050G} G{\"u}ver, T.,
{\"O}zel, F.\ 2009, \mnras, 400, 2050 

\bibitem[Hammer et al.(2007)]{2007ApJ...662..322H} Hammer, F., Puech, M., 
Chemin, L., Flores, H., \& Lehnert, M.~D.\ 2007, \apj, 662, 322 

\bibitem[Hickox \& Markevitch(2006)]{2006ApJ...645...95H} Hickox, R.~C., \&
Markevitch, M.\ 2006, \apj, 645, 95

\bibitem[Ho et al.(1997)]{1997ApJS..112..315H} Ho, L.~C., Filippenko,
A.~V., \& Sargent, W.~L.~W.\ 1997, \apjs, 112, 315

\bibitem[Hornschemeier et al. (2001)]{} Hornschemeier, A.E., et~al.\ 2001, ApJ,
554, 742

\bibitem[Hornschemeier et al.(2002)]{2002ApJ...568...82H} Hornschemeier,
A.~E., et al.\ 2002, \apj, 568, 82

\bibitem[Howell et al.(2010)]{2010arXiv1004.0985H} Howell, J.~H., et al.\ 
2010, \apj, in-press (arXiv:1004.0985) (H10)

\bibitem[Imanishi et al.(2007)]{2007ApJS..171...72I} Imanishi, M., Dudley, 
C.~C., Maiolino, R., Maloney, P.~R., Nakagawa, T., 
\& Risaliti, G.\ 2007, \apjs, 171, 72 

\bibitem[Imanishi et al.(2010)]{2010ApJ...709..801I} Imanishi, M., 
Maiolino, R., \& Nakagawa, T.\ 2010, \apj, 709, 801 

\bibitem[Isobe \& Feigelson(1990)]{1990BAAS...22..917I} Isobe, T., \&
Feigelson, E.\ 1990, \baas, 22, 917

\bibitem[Iwasawa et al.(2009)]{2009ApJ...695L.103I} Iwasawa, K., Sanders,
D.~B., Evans, A.~S., Mazzarella, J.~M., Armus, L., \& Surace, J.~A.\ 2009,
\apjl, 695, L103 (Iw09)

\bibitem[Jarrett et al.(2003)]{2003AJ....125..525J} Jarrett, T.~H., 
Chester, T., Cutri, R., Schneider, S.~E., 
\& Huchra, J.~P.\ 2003, \aj, 125, 525

\bibitem[Jenkins et al.(2005)]{2005MNRAS.357..109J} Jenkins, L.~P., 
Roberts, T.~P., Ward, M.~J., \& Zezas, A.\ 2005, \mnras, 357, 109 

\bibitem[Kaaret \& Alonso-Herrero(2008)]{2008ApJ...682.1020K} Kaaret, P., \&
Alonso-Herrero, A.\ 2008, \apj, 682, 1020 

\bibitem[Keel et al.(1985)]{1985AJ.....90..708K} Keel, W.~C., Kennicutt,
R.~C., Jr., Hummel, E., \& van der Hulst, J.~M.\ 1985, \aj, 90, 708

\bibitem[Kennicutt(1998)]{1998ARA&A..36..189K} Kennicutt, R.~C., Jr.\ 1998, 
\araa, 36, 189

\bibitem[Kennicutt et al.(2003)]{2003PASP..115..928K} Kennicutt, R.~C., 
Jr., et al.\ 2003, \pasp, 115, 928 

\bibitem[Kewley et al.(2001)]{2001ApJS..132...37K} Kewley, L.~J., Heisler,
C.~A., Dopita, M.~A., \& Lumsden, S.\ 2001, \apjs, 132, 37

\bibitem[Kim et al.(2007)]{2007ApJ...659...29K} Kim, M., Wilkes, B.~J., 
Kim, D.-W., Green, P.~J., Barkhouse, W.~A., Lee, M.~G., Silverman, J.~D., 
\& Tananbaum, H.~D.\ 2007, \apj, 659, 29 

\bibitem[Kroupa(2001)]{2001MNRAS.322..231K} Kroupa, P.\ 2001, \mnras, 322, 
231 

\bibitem[Lavalley et al.(1992)]{1992ASPC...25..245L} Lavalley, M., Isobe,
T., \& Feigelson, E.\ 1992, ASP Conf.~Ser.~ 25: Astronomical Data Analysis
Software and Systems I, 25, 245


\bibitem[Lehmer et al.(2008)]{2008} Lehmer, B.~D., et al.\ 2008, \apj, 681,
1163

\bibitem[Levenson et al.(2004)]{2004ApJ...602..135L} Levenson, N.~A., 
Weaver, K.~A., Heckman, T.~M., Awaki, H., 
\& Terashima, Y.\ 2004, \apj, 602, 135 

\bibitem[Luo et al.(2008)]{2008ApJS..179...19L} Luo, B., et al.\ 2008,
\apjs, 179, 19

\bibitem[McKee 
\& Williams(1997)]{1997ApJ...476..144M} McKee, C.~F., \& Williams, J.~P.\ 1997, \apj, 476, 144

\bibitem[Melnick 
\& Mirabel(1990)]{1990A&A...231L..19M} Melnick, J., \& Mirabel, I.~F.\ 1990, \aap, 231, L19 

\bibitem[Moretti et al.(2003)]{2003ApJ...588..696M} Moretti, A., Campana, 
S., Lazzati, D., \& Tagliaferri, G.\ 2003, \apj, 588, 696 

\bibitem[Ogle et al.(2003)]{2003A&A...402..849O} Ogle, P.~M., Brookings, T.,
Canizares, C.~R., Lee, J.~C., \& Marshall, H.~L.\ 2003, \aap, 402, 849 

\bibitem[Persic et al.(2004)]{2004A&A...419..849P} Persic, M., Rephaeli, Y.,
Braito, V., Cappi, M., Della Ceca, R., Franceschini, A., \& Gruber, D.~E.\
2004, \aap, 419, 849

\bibitem[Persic \& Rephaeli(2002)]{2002A&A...382..843P} Persic, M., \&
Rephaeli, Y.\ 2002, \aap, 382, 843

\bibitem[Persic \& Rephaeli(2007)]{2007A&A...463..481P} Persic, M., \&
Rephaeli, Y.\ 2007, \aap, 463, 481

\bibitem[Ptak et al.(2003)]{2003ApJ...592..782P} Ptak, A., Heckman, T., 
Levenson, N.~A., Weaver, K., \& Strickland, D.\ 2003, \apj, 592, 782 

\bibitem[Ranalli et 
al.(2003)]{2003A&A...399...39R} Ranalli, P., Comastri, A., \& Setti, G.\ 2003, \aap, 399, 39 

\bibitem[Risaliti et al.(2009)]{2009ApJ...696..160R} Risaliti, G., et al.\ 
2009a, \apj, 696, 160

\bibitem[Risaliti et al.(2009)]{2009ApJ...705L...1R} Risaliti, G., et al.\ 
2009b, \apjl, 705, L1 

\bibitem[Sandage \& Tammann(1981)]{1981RSA...C...0000S} Sandage, A., \&
Tammann, G. A. 1981, Revised Shapley-Ames Catalog of Bright Galaxies
(Publ.~635; Washington, DC: Carnegie Inst.)

\bibitem[Sanders \& Mirabel(1996)]{1996ARA&A..34..749S} Sanders, D.~B., \&
Mirabel, I.~F.\ 1996, \araa, 34, 74

\bibitem[Sanders et al.(2003)]{2003AJ....126.1607S} Sanders, D.~B., 
Mazzarella, J.~M., Kim, D.-C., Surace, J.~A., 
\& Soifer, B.~T.\ 2003, \aj, 126, 1607

\bibitem[Sanders et al.(1988)]{1988ApJ...328L..35S} Sanders, D.~B., Soifer, 
B.~T., Elias, J.~H., Neugebauer, G., \& Matthews, K.\ 1988, \apjl, 328, L35 

\bibitem[Scoville et al.(2000)]{2000AJ....119..991S} Scoville, N.~Z., et 
al.\ 2000, \aj, 119, 991 

\bibitem[Siebenmorgen \& Kr{\"u}gel(2007)]{2007A&A...461..445S} Siebenmorgen,
R., \& Kr{\"u}gel, E.\ 2007, \aap, 461, 445

\bibitem[Siebenmorgen et 
al.(2008)]{2008A&A...488...83S} Siebenmorgen, R., et al.\ 2008, \aap, 488, 83 

\bibitem[Smith 
\& Wilson(2003)]{2003ApJ...591..138S} Smith, D.~A., \& Wilson, A.~S.\ 2003, \apj, 591, 138

\bibitem[Soifer et al.(2000)]{2000AJ....119..509S} Soifer, B.~T., et al.\ 
2000, \aj, 119, 509 

\bibitem[Soria et al.(2009)]{2009ApJ...695.1614S} Soria, R., Risaliti, G., 
Elvis, M., Fabbiano, G., Bianchi, S., \& Kuncic, Z.\ 2009, \apj, 695, 1614 

\bibitem[Spergel et al.(2003)]{2003ApJS..148..175S} Spergel, D.~N., et al.\ 
2003, \apjs, 148, 175 

\bibitem[Strateva 
\& Komossa(2009)]{2009ApJ...692..443S} Strateva, I.~V., \& Komossa, S.\ 2009, \apj, 692, 443

\bibitem[Swartz et al.(2004)]{2004ApJS..154..519S} Swartz, D.~A., Ghosh, 
K.~K., Tennant, A.~F., \& Wu, K.\ 2004, \apjs, 154, 519 

\bibitem[Veilleux et al.(1995)]{1995ApJS...98..171V} Veilleux, S.,
Kim, D.-C., Sanders, D.~B., Mazzarella, J.~M., \& Soifer, B.~T.\ 1995,
\apjs, 98, 171

\bibitem[Veron-Cetty \& Veron(1986)]{1986A&AS...66..335V} Veron-Cetty,
M.-P., \& Veron, P.\ 1986, \aaps, 66, 335

\bibitem[Wang et al.(2009)]{2009ApJ...694..718W} Wang, J., Fabbiano, G., 
Elvis, M., Risaliti, G., Mazzarella, J.~M., Howell, J.~H., 
\& Lord, S.\ 2009, \apj, 694, 718

\bibitem[Young et al.(2001)]{2001ApJ...556....6Y} Young, A.~J., Wilson, 
A.~S., \& Shopbell, P.~L.\ 2001, \apj, 556, 6

\bibitem[Yuan et al.(2010)]{2010ApJ...709..884Y} Yuan, T.-T., Kewley,
L.~J., \& Sanders, D.~B.\ 2010, \apj, 709, 884

\bibitem[Zezas et al.(2003)]{2003ApJ...594L..31Z} Zezas, A., Ward, M.~J., 
\& Murray, S.~S.\ 2003, \apjl, 594, L31 

%
%%%%%%%%%%%%%%%%%%%%%%%%%%%%%%%%%%%%%%%%%%%%%%%%%%%%%%%%%%%%%%%%%%%%%%%%%%%%%%%%%%
\end{thebibliography}
\end{document}